# Radiation-driven winds of hot luminous stars

## XVII. Parameters of selected central stars of PN from consistent optical and UV spectral analysis and the universality of the mass–luminosity relation

C. B. Kaschinski, A. W. A. Pauldrach, and T. L. Hoffmann

Institut für Astronomie und Astrophysik der Universität München, Scheinerstraße 1, 81679 München, Germany
e-mail: corni@usm.lmu.de, e-mail: uh10107@usm.lmu.de, e-mail: hoffmann@usm.lmu.de

**ABSTRACT**

*Context.* The commonly accepted mass–luminosity relation of central stars of planetary nebulae (CSPNs) might not be universally valid. While the atmospheric parameters $T_{\rm eff}$, $\log g$, and the He abundances have in the past been determined using fits to photospheric H and He absorption lines from plane-parallel model atmospheres, the masses and luminosities could not be derived independently this way, and were instead taken from theoretical evolutionary models. Improvements in hydrodynamically selfconsistent modelling of the stellar atmospheric outflows now allow using fits to the wind-sensitive features in the UV spectra to consistently determine the stellar radii, masses, and luminosities without assuming a mass–luminosity relation. Recent application to a sample of CSPNs in an earlier paper of this series raised questions regarding the validity of the mass–luminosity relation of central stars of planetary nebulae.
*Aims.* The results of the earlier UV analysis are reassessed by means of a simultaneous comparison of both the observed optical and UV spectra with corresponding synthetic spectra. The synthetic optical and UV spectra are computed simultaneously from the same atmospheric models, using the same model atmosphere code. Synthetic spectra for the two central stars NGC 6826 and NGC 2392 are computed using parameter sets from two different published analyses to check their compatibility to the observations.
*Methods.* Using the different published stellar parameter sets, derived on the one hand by a consistent UV analysis, and on the other hand from fits to optical H and He lines, we calculate corresponding optical and UV spectra with our model atmosphere code. We have improved this model atmosphere code by implementing Stark broadening for hydrogen and helium lines, thus allowing us to obtain consistent optical H and He line profiles simultaneously with our state-of-the-art modelling of the UV-spectrum.
*Results.* Optical line profiles computed with the consistent parameter sets from the UV analysis yield line profiles with good agreement to the observations (with small discrepancies in the emission lines corresponding to about a factor of one half in the mass loss rate). Spectra computed with the stellar parameter sets from the optical analysis in the literature and corresponding *consistent* wind parameters, however, show large discrepancies to the observed spectra, in the optical as well as in the UV. We conclude that the published optical analyses give good fits to the observed spectrum only because the wind parameters assumed in these analyses are inconsistent to their stellar parameters. By enforcing consistency between stellar and wind parameters, stellar parameters are obtained which disagree with the core-mass–luminosity relation for the two objects analyzed. This disagreement is also evident from a completely different approach: an investigation of the dynamical wind parameters.

**Key words.** stars: central stars of planetary nebulae – atmospheres – winds, outflows – evolution – fundamental parameters

## 1. Introduction

In recent years there has been substantial progress in the modelling of expanding atmospheres of hot stars. Current state-of-the-art wind models dealing with homogeneous, stationary, spherically symmetric, radiatively driven, extended, outflowing atmospheres can now produce synthetic UV spectra of O stars that resemble the observed ones nearly perfectly. A complete model atmosphere calculation of this kind involves solving both the hydrodynamics and the so-called non-LTE problem,[1] comprising a simultaneous solution of the total interdependent system of the radiative transfer, the rate equations for all important elements, and the energy equation. The primary diagnostic output of such a computation is a predicted, or synthetic, spectrum, which can be compared to an observed UV spectrum. The fundamental stellar parameters are determined by varying the model parameters until a match to the observed spectrum is achieved. Using this new generation of realistic stellar model atmospheres, Pauldrach et al. (2001) and Pauldrach (2003) have already presented an analysis of the massive O supergiants HD 30614 ($\alpha$ Cam) and HD 66811 ($\zeta$ Pup) that provided good matches to the observable UV spectra, thereby determining the basic stellar parameter sets of these objects.

By solving the stationary hydrodynamic equations (in which the acceleration driving the outflow depends in turn on the occupation numbers and the radiation field) simultaneously with the non-LTE problem, a solution is obtained that is hydrodynamically consistent and provides the velocity law as well as the mass-loss rate, two key quantities determining the appearance of the UV spectrum. (For details of the physical and technical background see Pauldrach 1987, Pauldrach et al. 1994, and details in Pauldrach et al. 2004.) Such a consistent treatment has very important consequences for the analysis of the UV spectral features, since it provides information about the complete set of the basic stellar parameters: the effective temperature $T_{\rm eff}$, the radius $R$ (or equivalently, the luminosity $L$), the mass $M$, the terminal wind velocity $v_\infty$, and the mass loss rate $\dot{M}$. Thus, a purely spectroscopic method allowing the determination of $L$ and $M$ exists.

[1] Due to the high radiation intensities and the low densities in hot star atmospheres, the ionic occupation numbers can deviate strongly from local thermodynamic equilibrium (LTE), and thus a much more general treatment is required. Our approach is described in Pauldrach et al. 2001.

Pauldrach et al. (2004) have applied this method to O-type central stars of planetary nebulae, thus providing, for the first time, an *independent* test of the predictions from post-AGB evolutionary calculations (see, for instance, Blöcker 1995). This was not possible in the earlier work on CSPNs, which was based on plane-parallel, non-LTE model atmospheres (e.g., Mendez et al. 1988a, Mendez et al. 1988b). Lacking a consistent solution of the hydrodynamics of the atmospheric outflow, the corresponding model fits to optical hydrogen and helium photospheric absorption lines could only provide information about surface temperature, helium abundance, and surface gravity ($\log g$), whereas stellar masses or luminosities could not be derived from the spectra, and additional observational data was needed to furnish information about the physical size of the stars, such as their distances. But for most CSPNs, unfortunately, reliable distance measurements are not available.

In the earlier work, therefore, the masses and luminosities of these objects were obtained from the positions of the CSPNs in the $\log g$–$\log T_{\rm eff}$ diagram (both the effective temperatures and the surface gravities being available from spectroscopic analysis) by comparing these with theoretical post-AGB tracks for given masses. This procedure to determine the CSPN masses is thus based entirely on the assumption that the evolutionary models give the correct relation between stellar mass and luminosity.

Kudritzki et al. (1997) performed such an analysis for a sample of 9 selected O-type CSPNs, additionally modeling the H$\alpha$ line profiles to determine the mass loss rates, and found the O-type CSPNs to lie along the wind-momentum–luminosity relation defined by the massive O stars, albeit with a somewhat larger scatter. (The tight correlation between the so-called modified wind-momentum rate $D = \dot{M} v_\infty \sqrt{R}$ and the stellar luminosity $L$ of massive O stars was found empirically by Kudritzki et al. (1995) and was subsequently explained by Puls et al. (1996) using the theory of radiatively driven winds.) This was a further indication that the winds of O-type CSPNs are radiatively driven and that the atmospheres of massive O stars and O-type CSPNs are governed by the same physics, confirming the work of Pauldrach et al. (1988). Nevertheless, one of the surprising results of Kudritzki et al. (1997) was the large fraction of high-mass CSPNs in their sample.

In another study, Napiwotzki et al. (1999) determined masses for a sample of 46 hot DA white dwarfs selected from the Extreme UV Explorer (EUVE) and the ROSAT Wide Field Camera bright source lists. They found a peak mass of 0.59 $M_\odot$, in agreement with many other studies, but also found a non-negligible fraction of white dwarfs with masses in excess of 1 $M_\odot$.

These surprising results prompted Tinkler & Lamers (2002) to check the consistency of stellar and wind parameters for a larger sample of CSPNs. As a result of scaling the distances and stellar parameters according to their method, however, they obtained no clear dependence of wind momentum on luminosity. This brought up a conflict between the predictions of post-AGB evolutionary theory and the theory of radiatively driven stellar winds.

The work of Pauldrach et al. (2004) compounded the situation further. The masses obtained from hydrodynamically consistent modelling of the UV spectra of the CSPNs of the particular sample that had already been analyzed by Kudritzki et al. (1997) were for the most part even larger than those determined by Kudritzki et al. Figure 1 shows that the luminosities derived from the UV analysis lie within the range expected from their prominent P-Cygni spectral features, but a much larger spread in the masses was obtained, up to the critical Chandrasekhar mass limit for white dwarfs.

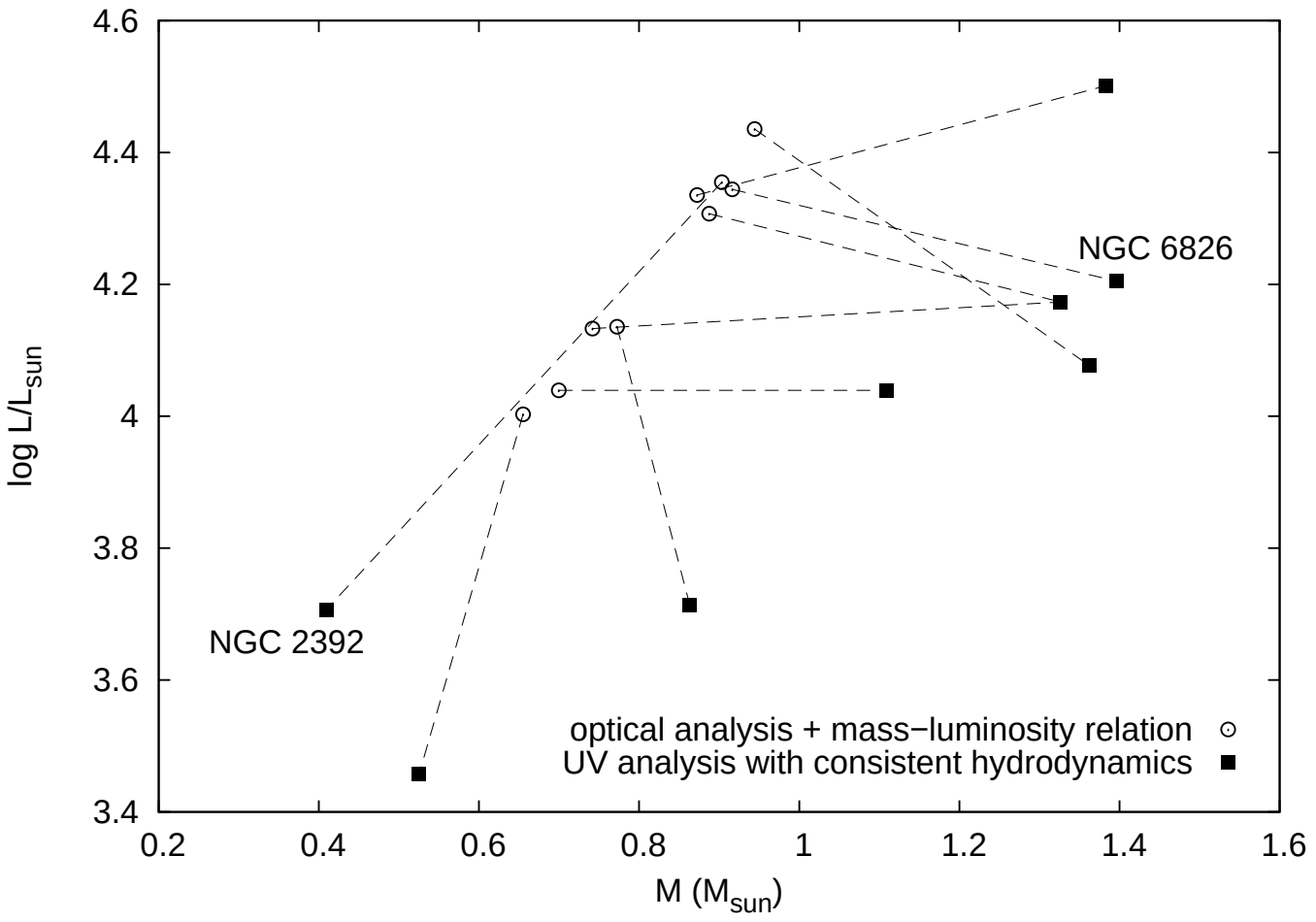

**Fig. 1.** Luminosities vs. masses derived for a sample of CSPNs using a hydrodynamically consistent UV spectral analysis (filled squares; Pauldrach et al. 2004) vs. those obtained from an analysis of optical hydrogen and helium lines and assuming a mass–luminosity relation from theoretical post-AGB evolutionary tracks (open circles; Kudritzki et al. 1997). Although the luminosities derived with the UV analysis lie in the expected range, the obtained masses ($0.4\,M_\odot$ up to $1.4\,M_\odot$) deviate severely from the masses for the evolutionary tracks.

This result was even more disturbing for the community. To clarify the situation, Napiwotzki (2006) investigated the kinematical properties of the sample stars of Kudritzki et al. (1997) and Pauldrach et al. (2004). He found NGC 6826, one of the sample stars, to be indeed close to the old thin disk, but not the young thin disk as Napiwotzki suspected. Apart from one star he found better – but still unsatisfactory – agreement with respect to the kinematical properties of his sample by arbitrarily assuming a mass of 0.565 $M_\odot$ for *all* CSPNs, ignoring the fact that Kudritzki et al. (1997), based on the core-mass–luminosity relation *and* well-established spectral analysis techniques, had found masses almost a factor of two larger for most objects of the sample. As these latter masses are not based on an arbitrary assumption, but on a sophisticated spectroscopic analysis, these masses are certainly to be preferred to those assumed by Napiwotzki. Although Napiwotzki did not show results for the Kudritzki et al. masses, interpolating between the positions for the Pauldrach et al. (2004) masses and for 0.565 $M_\odot$ in Napiwotzki's diagrams, we see that the masses of Kudritzki et al. also contradict the kinematical positions. Thus, the kinematic evolution of the selected group of CSPNs (which had been chosen for their pronounced wind spectra and their uncharacteristically very high luminosities) is obviously not as straightforward as one might think on basis of simple arguments. This clearly means that the kinematics of CSPN, at least for a statistically not representative sample such as this one, are much less reliable than the well-understood behavior of stellar atmospheres.[2]

[2] Moreover, Napiwotzki employed a circular argument by basing the primary assumption – the theoretical age of the sample – on the theory which has been shown not to work for the sub-sample of objects, in order to show that that theory cannot be wrong.

Although the result of a much larger spread in the masses, up to the critical mass limit for white dwarfs, of the selected group of CSPNs might be of relevance for the controversially discussed precursor scenarios of type Ia supernovae (cf. Pauldrach 2005), the discrepancy to the optical analysis of Kudritzki et al. (1997) is *still* surprising, since the same model atmospheres worked perfectly well for massive O stars, using exactly the same physics (cf. Pauldrach et al. 2011).

All this is a strong hint that there is something fundamental we do not yet understand about the formation of CSPNs. In this paper we try to provide further constraints to clarify the situation. We do this by testing whether, if both optical and UV analysis are based on an adequate consistent treatment of the expanding CSPN atmospheres, analyses from optical spectra necessarily yield different stellar parameters than analyses from UV-spectra, or whether the discrepancies between the analyses result as a consequence of the missing consistency between the stellar and the assumed wind parameters in the published optical analyses of Kudritzki et al. 1997 and Kudritzki et al. 2006.

In the following we will first outline the two different modeling techniques for deriving the stellar parameters (Section 2) and describe the implementation of the Stark broadening tables in our stellar atmosphere code and the test calculations performed using H and He lines (Section 3). After briefly identifying the optical and UV observations we have used in this study (Section 4) we describe the model runs performed using the published parameter sets for the two CSPNs selected from the sample previously studied by Pauldrach et al. (2004) and Kudritzki et al. (1997), NGC 6826 and NGC 2392, and discuss the resulting synthetic optical and UV spectra in comparison to the observed spectra (Section 5).

With regard to the stellar parameters – $R$, $L$ and $M$ – of a sample of CSPN stars with pronounced wind features we include in this section a discussion of the dynamical parameters as an additional point. This discussion should be considered as an extension to the investigation of Pauldrach et al. (1988) who showed that the calculated terminal wind velocities are in agreement with the observations and therefore allow an independent determination of stellar masses and radii. With respect to this result, different sets of stellar masses and radii applied to our sample of stars should therefore lead at least partly to an inconsistent behavior with regard to predictions of the radiation-driven wind theory.[3] Here, such an inconsistent behavior is realized by the ratios of the terminal wind velocities $v_\infty$ and the escape velocities $v_{esc}$ of the stars (cf. Pauldrach et al. 1988 and Pauldrach et al. 1990). We compare our results of the $v_\infty/v_{esc}$ obtained with our improved models for the CSPN sample not just to the corresponding "observed ratios", which are based on the the mass–luminosity relation of CSPN, but also to the ratios of an O star sample and its corresponding observations. We interpret and summarize our results in Section 6.

## 2. Methods

### 2.1. Parameter determination using hydrodynamic models and the UV spectrum

The winds of O-type stars are driven by radiative absorption in spectral lines, a circumstance reflected in the existence of the wind-momentum–luminosity relation. This means that the velocity and strength of the wind are not free parameters, but instead explicit functions of the stellar parameters and the atmospheric chemical composition. This in turn implies that if the dependence is known or can be computed, then a measurement of the density and velocity structure of the atmospheric outflow will also yield the fundamental stellar parameters.

The UV spectrum between 1000 and 2000 Å is well suited for this measurement. It contains P-Cygni-type profiles of resonance lines of several ions of C, N, O, Si, S, P, which remain strong until far out in the wind, as well as hundreds of strongly wind-contaminated lines of Fe IV, Fe V, Fe VI, Cr V, Ni IV, Ar V, and Ar VI, formed in a large range of depths from deep within the photosphere to the transition region into the "wind". But in order to extract the information about the abundances and wind parameters, and from these the stellar parameters, a sophisticated analysis is required.

Our method for modeling the atmosphere and computing the emergent UV spectrum in order to deduce the fundamental stellar parameters via a comparison with the observed spectrum is based on the premise that the winds are radiation-driven, homogeneous, stationary, and spherically symmetric. It incorporates a consistent treatment of the blocking and blanketing influence of all metal lines in the entire sub- and supersonically regions and a full non-LTE treatment of all level populations. Although a detailed description of the procedure is given by Pauldrach et al. (2001), Pauldrach (2003), Pauldrach et al. (2004), and Pauldrach et al. (2011), we will give here an overview of the physics to be treated.

The general concept. Although the basis of our approach in constructing detailed atmospheric models for hot stars is the concept of homogeneous, stationary, and spherically symmetric radiation-driven atmospheres, the method is not simple, since modeling the atmospheres of hot stars involves the replication of a tightly interwoven mesh of physical processes: the equations of radiation hydrodynamics including the energy equation, the statistical equilibrium for all important ions with detailed atomic physics (a detailed description of our atomic models is to be found in Sect. 3 and Table 1 of Pauldrach et al. (2001) and in Sect. 2 of Pauldrach et al. (1994) where several Tables and Figures illustrating and explaining the overall procedure are shown), and the radiative transfer equation at all transition frequencies have to be solved simultaneously.

The principal features of the method are:

- *The hydrodynamic equations* are solved. Here the crucial term is the radiative acceleration with minor contributions from continuous absorption and major contributions from scattering and line absorption (including the effects of line-overlap and multiple scattering, cf. Fig. 2). We note that the consistent treatment of the hydrodynamics is a crucial point, because the hydrodynamics affects the non-LTE model via the density structure and the velocity field, and the radiative transfer with respect to Doppler-shifted spectral lines, but in turn is controlled by the line force determined by the occupation numbers and the radiative transfer of the non-LTE model. Back reaction mechanisms are therefore inherently involved in the procedure.
- *The occupation numbers* are determined by the *rate equations* containing collisional and radiative transition rates. Low-temperature dielectronic recombination is included and Auger ionization due to K-shell absorption (essential for C, N, O, Ne, Mg, Si, and S) of soft X-ray radiation arising from shock-heated matter is taken into account using detailed atomic models for all important ions. Note that the hydrodynamical equations are coupled directly with the rate equations. The velocity field enters into the radiative rates via the Doppler shift, while the density is important for the collisional rates and the total number density.

[3] Note that based on striking observational properties of CSPN winds Pauldrach et al. (1988) where able to show that the winds of CSPN are driven by radiation pressure.

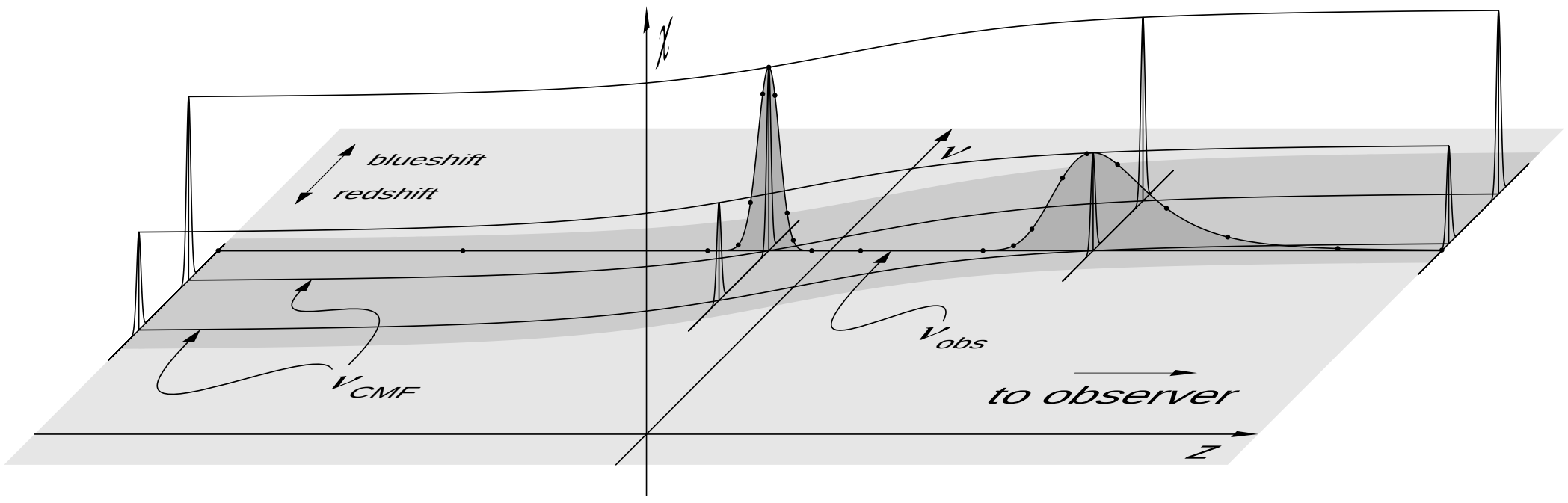

**Fig. 2.** Diagram illustrating the basic relationship of the rest-frame frequencies of spectral lines ($\nu_{\rm CMF}$) to observer's frame frequency ($\nu_{\rm obs}$) for one particular (non-core) $p$-ray in the spherically symmetric geometry ($p, z$ geometry). Shown are two spectral lines of different opacity $\chi$ which get shifted across the observer's frame frequency by the velocity field in the wind. The dots represent the stepping points of the adaptive microgrid used in solving the transfer equation in the radiative line transfer. The method employed is an integral formulation of the transfer equation using an adaptive stepping technique on every $p$-ray in which the radiation transfer in each micro-interval is solved as a weighted sum on the microgrid (cf. Pauldrach et al. 2001): $I(\tau_0(p,z)) = I(\tau_n)e^{-(\tau_n-\tau_0)} + \sum_{i=0}^{n-1}\left(e^{-(\tau_i-\tau_0)}\int_{\tau_i}^{\tau_{i+1}} S(\tau)e^{-(\tau-\tau_i)}\,{\rm d}\tau(p,z)\right)$ where $I$ is the specific intensity, $S$ is the source function and $\tau$ is the optical depth (increasing from $\tau_0$ on the right to $\tau_n$ on the left in the figure). To accurately account for the variation of the line opacities and emissivities due to the Doppler shift, all line profile functions are evaluated correctly for the current microgrid-($z, p$)-coordinate on the ray, thus effectively resolving individual line profiles. Based on that, application of the Sobolev technique gives for the radiative line acceleration (cf. Pauldrach et al. 2011): $g_{\rm lines}(r) = \frac{2\pi}{c}\frac{1}{\rho(r)}\sum_{\rm lines}\chi_{\rm line}(r)\int_{-1}^{+1} I_{\nu_0}(r,\mu)\frac{1-e^{-\tau_s(r,\mu)}}{\tau_s(r,\mu)}\mu\,{\rm d}\mu$ where $\tau_s(r,\mu) = \chi_{\rm line}(r)\frac{c}{\nu_0}\left[(1-\mu^2)\frac{v(r)}{r} + \mu^2\frac{{\rm d}v(r)}{{\rm d}r}\right]^{-1}$ is the Sobolev optical depth, and $\nu_0$ is the frequency at the center of each line – thus, the effects of line-overlap and multiple scattering are naturally included ($\chi_{\rm line}(r)$ is the local line absorption coefficient, $\mu$ is the cosine of the angle between the ray direction and the outward normal on the spherical surface element, and $c$ is the speed of light).[4]

- *The spherical transfer equation* which yields the radiation field at every depth point, including the thermalized layers where the diffusion approximation is used as inner boundary, is solved for the total opacities and source functions of all important ions. Hence, the influence of the spectral lines – the strong EUV *line blocking* including the effects of line-overlap (cf. Fig. 2) – which affects the ionizing flux that determines the ionization and excitation of levels, is naturally taken into account. This is also the case for the effect of Stark-broadening, which is essential for the diagnostic use of certain spectral lines. *Stark-broadening* has therefore, as a new feature, been included in our procedure (cf. Sect. 3). Moreover, the *shock source functions* produced by radiative cooling zones which originate from a simulation of shock heated matter, together with K-shell absorption, are also included in the radiative transfer (the shock source function is incorporated on the basis of an approximate calculation of the volume emission coefficient of the X-ray plasma in dependence of the velocity-dependent post-shock temperatures and a filling factor).

- *The temperature structure* is determined by the microscopic *energy equation* which, in principle, states that the luminosity must be conserved in the atmosphere. *Line blanketing* effects which reflect the influence of line blocking on the temperature structure are taken into account.

The iterative solution of the total system of equations yields the hydrodynamic structure of the wind (i.e., the *mass-loss rate* and the *velocity structure*) together with *synthetic spectra* and *ionizing fluxes*.

The analysis technique. In general, the analysis technique involves the following steps: first, a preliminary inspection of the UV and/or visual spectrum of the star gives a guess for $T_{\rm eff}$, and, together with a reasonable estimate of the mass $M$ and radius $R$ of the star, we use this to compute an initial atmospheric model and its emergent UV spectrum. Since the ionization balance in the wind depends strongly on the strength of the radiation field and thus on the effective temperature, we can then refine our value for $T_{\rm eff}$ by comparing the strengths of the lines of successive ionization stages of several elements.[5] Given the current

[4] Note that a comparison of the line acceleration of strong and weak lines evaluated with the comoving frame method and the Sobolev technique without consideration of the continuum acceleration is presented in Fig. 5 of Pauldrach et al. (1986), and with the comoving frame method and the Sobolev-with-continuum technique with consideration of the complete continuum acceleration in Fig. 3 of Puls & Hummer (1988), showing the excellent agreement of the two methods. It is thus important to realize that the accuracy of the calculation of the radiative acceleration is of the same quality as that of the synthetic spectrum, since the radiative acceleration is calculated analogously and in parallel to the synthetic spectrum. This means that the velocity field $v(r)$ and the mass loss rate $\dot{M}$, which are just functions of the basic stellar parameters and the radiative acceleration, are as realistic as the synthetic spectrum is. Finally, the hydrodynamics is solved by iterating the complete continuum acceleration $g_{\rm cont}(r)$ (which includes in our case the force of Thomson scattering and of the continuum opacities $\chi_\nu^{\rm cont}(r)$ – free-free and bound-free – of all important ions (cf. Pauldrach et al. 2001)) together with the line acceleration $g_{\rm lines}(r)$ – obtained from the spherical NLTE model – and the density $\rho(r)$, the velocity $v(r)$, and the temperature structure $T(r)$ (cf. Pauldrach et al. 2011 and references therein).

[5] Especially for effective temperatures in the range of 30 000 to 40 000 K we have found the Fe IV/Fe V ionization balance well suited for this purpose. The wavelength range from 1400 to 1550 Å is dominated by lines from Fe V, whereas Fe IV lines dominate in the range from 1550 to 1650 Å, and a comparison of the relative strengths of the

set of stellar parameters, a consistent calculation of the hydrodynamics of the outflow yields the terminal velocity $\upsilon_\infty$ of the wind, measurable from the blue edge of the P-Cygni profiles of strong resonance lines in the observed UV spectrum, and the mass loss rate $\dot{M}$, reflected mainly in the overall strength of the lines in the transition region. In case the calculated $\upsilon_\infty$ of the model differs from the observed value, we need to modify the stellar mass until agreement is reached (since $\upsilon_\infty$ depends sensitively on $(M/R)^{1/2}$ according to the theory of radiation-driven winds). If the overall fit of the spectrum is not satisfactory we must modify the mass loss rate $\dot{M}$ via a change of $R$ (since $\log \dot{M} \sim \log L$, according to radiation-driven wind theory). The change in $R$ forces us to change the mass, too, in order to keep $\upsilon_\infty$ consistent with the observed value. (Additionally, we may correct the temperature slightly, if this improves the fit.) The new model is calculated and the process is repeated until we obtain a good simultaneous fit to all features in the observed spectrum. For a detailed description of the procedure used to derive all relevant parameters see Pauldrach et al. (2004) and Pauldrach et al. (2011).

### 2.2. Parameter determination using optical H and He lines

An alternative method which has been applied to the analysis of hot stars by other workers in the field is based on modelling the optical spectrum, in particular using the stellar H$\alpha$ profile to determine the mass loss rate (see, for instance, Puls et al. 1996 and references therein): the strength of H$\alpha$, if in emission, reacts very sensitively to changes in the density (being a recombination line), and thus the mass loss rate. The outflow velocity is usually assumed to be a function of the radius as a so-called "beta velocity law," $\upsilon(r) = \upsilon_\infty (1 - R/r)^\beta$, where the shape of the velocity field is dependent on the arbitrary parameter $\beta$ (the best-fit value of $\beta$ to be determined as part of the analysis).

The fitting procedure is usually as follows (see also Kudritzki et al. 1997): by using fits to the optical hydrogen and helium stellar absorption lines with theoretical profiles computed from plane-parallel non-LTE models, preliminary estimates of $T_{\rm eff}$, the surface gravity $\log g$, and the He abundance $Y_{\rm He}$ are made. To calculate the H$\alpha$ profile an estimate for the stellar mass $M$ as well as for the terminal velocity $\upsilon_\infty$ is needed in addition to the already estimated preliminary atmospheric parameters $T_{\rm eff}$, $\log g$, and $Y_{\rm He}$: the mass is derived from evolutionary tracks plotted in the $\log g$–$\log T_{\rm eff}$ diagram, assuming these to be correct, whereas $\upsilon_\infty$ is measured directly from the bluest edge of the strongest resonance line in the observed UV spectrum. To fit the stellar H$\alpha$ profile, different values for the parameter $\beta$ and the mass loss rate are used. Having a first estimate for the mass loss rate, new "unified" models (comprising photosphere and wind, parametrizing the outflow velocity law with the fit parameter $\beta$) are calculated to obtain improved theoretical profiles for the other diagnostic lines. The procedure is iterated until adequate fits to most of the stellar absorption and emission lines are achieved. It must be stressed that in this procedure it is only possible to obtain values for the stellar mass and the luminosity by assuming that the evolutionary models give us the correct relation between the two.

lines in these wavelength ranges in the observed and the synthetic UV spectrum usually allows the effective temperature to be constrained to within ±1000 K (see, for instance, Pauldrach et al. 2001).

**Table 1.** Parameters of the two test models used for comparing the resulting profile shapes of optical H and He lines computed with the two different stellar atmosphere codes WM-basic and Fastwind.

| Model | $T_{\rm eff}$ (K) | $\log g$ | $R$ ($R_\odot$) | $\dot{M}$ ($M_\odot$/yr) | $\upsilon_\infty$ (km/s) |
|---|---|---|---|---|---|
| D30 | 30000 | 3.85 | 12 | $8 \times 10^{-9}$ | 1800 |
| D45 | 45000 | 3.9 | 12 | $1.3 \times 10^{-6}$ | 3000 |

### 2.3. Combined analysis

Given the differences between the parameter sets derived from UV vs. optical analyses published in the literature, the obvious question was whether there is an intrinsic discrepancy between modelling the UV and the optical spectra. Up to now we were unable to answer this question since our model atmosphere code did not compute the shapes of the optical lines correctly, due to the fact that Stark broadening was not included in the calculations. (This is quite unimportant for the atmospheric structure and the UV spectra, but crucial if one wants to determine atmospheric parameters from the optical lines.) With the inclusion of Stark broadening (see below) we can now compute UV and optical spectra consistently from the same atmospheric model. Having this tool at hand, we will in this paper apply it to the parameter sets for two CSPNs published by Pauldrach et al. (2004) (UV analysis) and Kudritzki et al. (1997) (optical analysis), and compare the results to the observations.

## 3. Implementation of Stark broadening in the model atmosphere code WM-basic

Depending on the physical environment, the shapes of line profiles are affected by different physical processes. Natural broadening, for instance, dominates the shapes of the profiles at large frequency distances from the line centers. Doppler broadening, on the other hand, is always relevant and dominates the shapes close to the line center. In dense atmospheres the line shapes are also strongly influenced by interactions of radiating atoms or ions with the surrounding particles. This behavior reflects the so called pressure broadening and, as electric fields are involved, this type of broadening is also called Stark broadening. In order to include this effect in our model atmosphere code,[6] reliable profile functions for H, He I, and He II were required. For the H lines we used the data sets published by Vidal et al. (1973). Those for the He I lines have been implemented according to Griem (1964), Barnard et al. (1969) and Shamey (1964), whereas the data sets for the He II lines have been applied according to Schoening & Butler (1989).

To validate our implementation of Stark broadening, we have performed test calculations and compared our results to those obtained with the model atmosphere code Fastwind (Puls et al. 1996, Santolaya-Rey et al. 1997, Puls et al. 2005), a well-established code for modelling the optical spectra of hot stars.

[6] Since Stark broadening affects primarily the line wings, while the line cores are still dominated by Doppler broadening, as was shown by Herrero (1987) the inclusion of Stark broadening in the rate equations is of minor importance. (Herrero showed that the consideration of Stark broadening in both the rate equations and in the formal integral leads to results which are almost identical with results obtained for the case where Stark broadening is included just in the solution of the formal integral.) Hence, it is a common procedure to include Stark broadening only in the formal integral (cf. Santolaya-Rey et al. 1997).

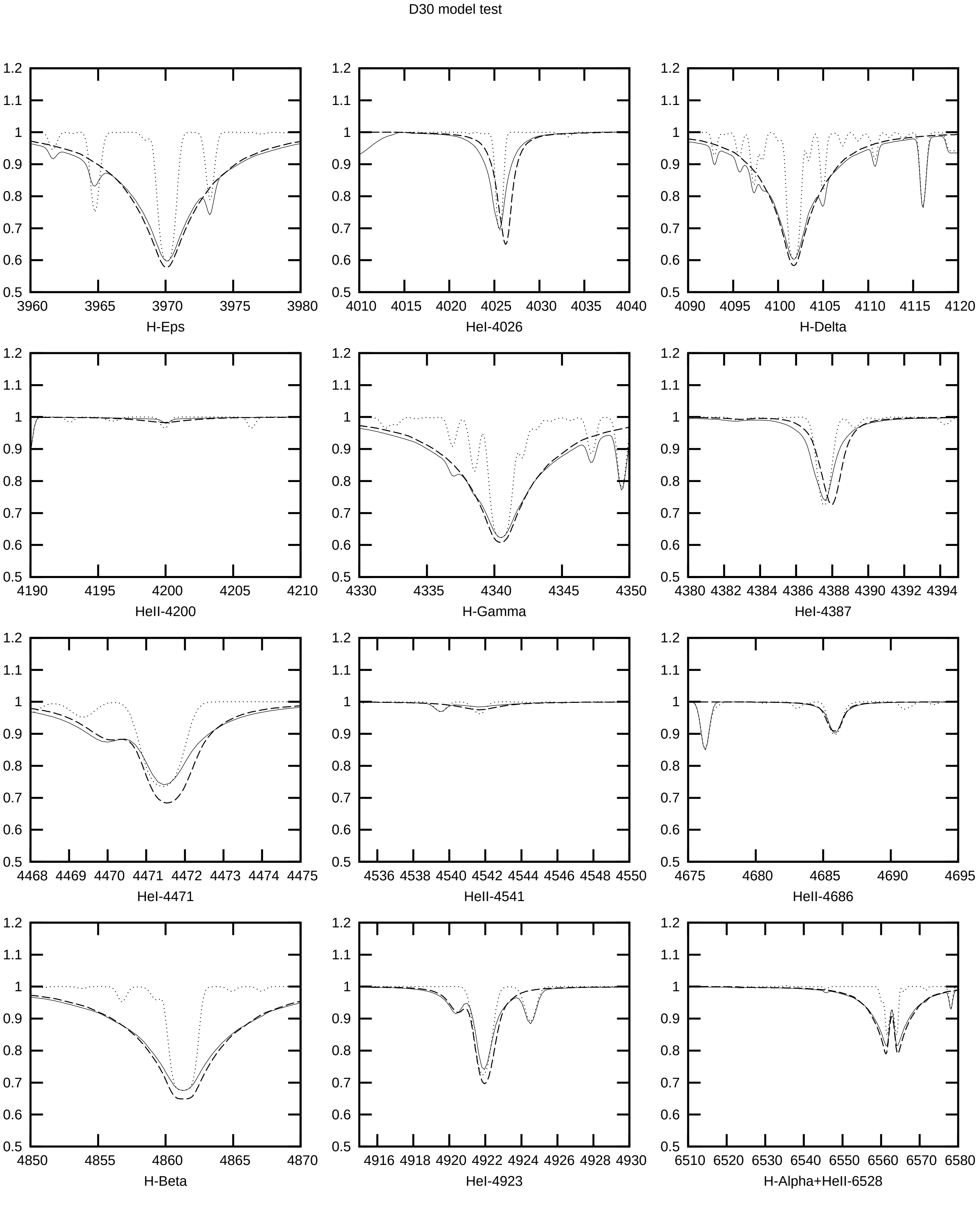

**Fig. 5.** H and He lines of the D30 model calculated with our stellar atmosphere code WM-basic (solid lines) compared to the H and He lines calculated with the model atmosphere code Fastwind (dashed lines). (The WM-basic model also contains additional metal lines not included in the Fastwind model.) Note the big difference between the H and He line profiles of the former WM-basic code (dotted lines) missing the Stark broadening to the present profiles (solid lines). Our improved models match the reference model lines almost perfectly. Small deviations due to differences in density and occupation numbers of the two codes are to expected.

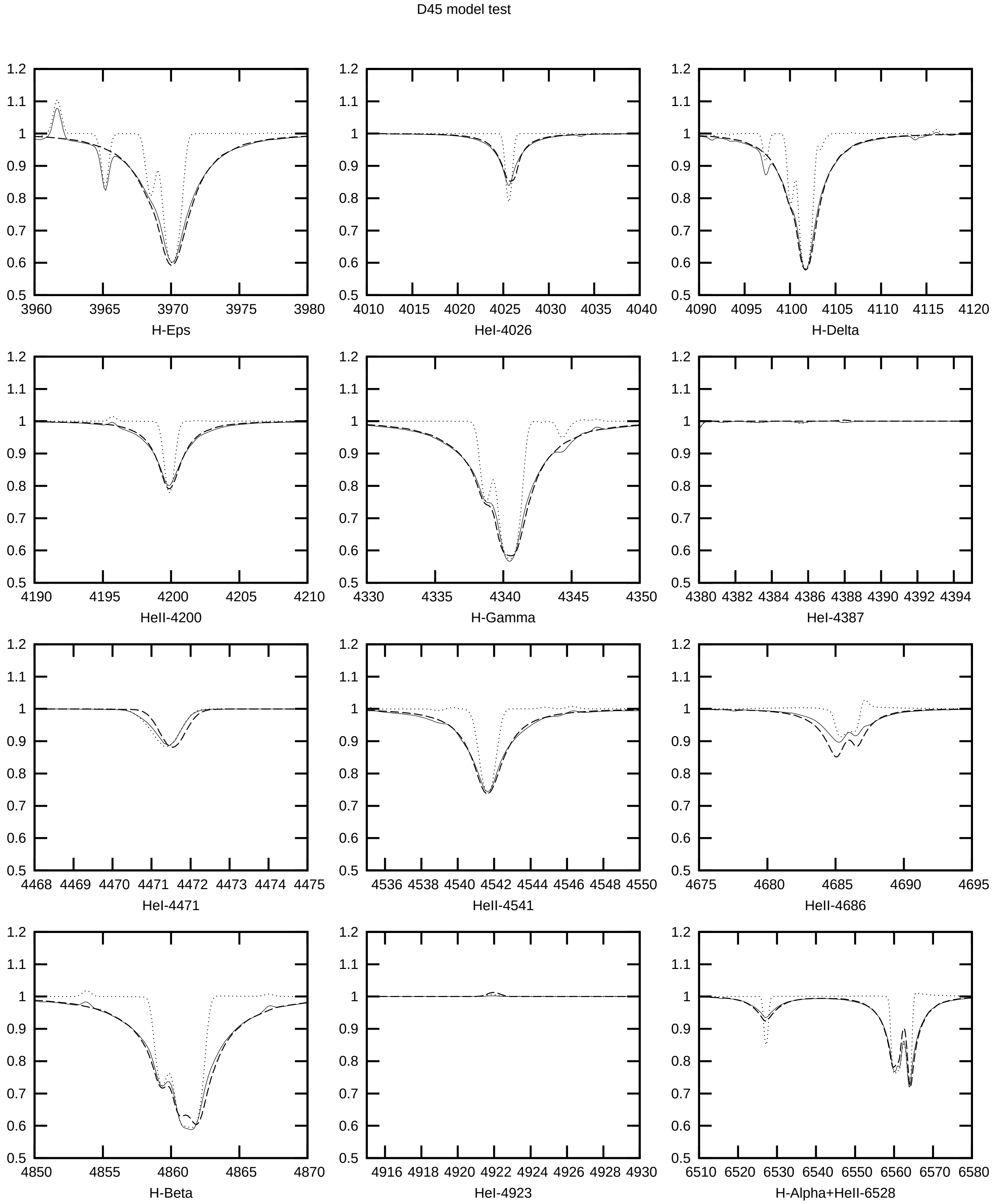

**Fig. 6.** H and He lines of the D45 model calculated with our stellar atmosphere code WM-basic (solid lines) compared to the H and He lines calculated with the model atmosphere code Fastwind (dashed lines). Note the big difference between the H and He line profiles of the former WM-basic code (dotted lines) missing the Stark broadening to the present profiles (solid lines). Again our improved models match the reference model lines almost perfectly. Small deviations due to differences in density and occupation numbers of the two codes are to expected.

**Table 2.** Stellar and wind parameters of the models used to compute the synthetic spectra presented in this work for the two CSPNs NGC 6826 and NGC 2392. P04 refers to the parameters derived by Pauldrach et al. (2004) from an analysis of the UV spectra, K97 refers to the optical analysis of Kudritzki et al. 1997, and K06 refers to the optical analysis of Kudritzki et al. 2006 (a dash in the "matches observation?" column indicates that a comparison to observations was not shown in the paper). "Consistent" means that the wind parameters are consistent with the stellar parameters as determined by our hydrodynamic calculations of the radiative driving force. We note that Kudritzki et al. 2006 obtained a much lower mass loss rate for NGC 6826 than Kudritzki et al. 1997.

| $T_{\rm eff}$ (K) | $R$ ($R_\odot$) | $\log L$ ($L_\odot$) | $M$ ($M_\odot$) | $\log g$ (cm/s$^2$) | $\dot{M}$ ($10^{-6} M_\odot$/yr) | $v_\infty$ (km/s) | parameter source stellar | wind | consistent? | matches observation? |
|---|---|---|---|---|---|---|---|---|---|---|
| | | | | | NGC 6826 | | | | | |
| 50000 | 2.0 | 4.4 | 0.92 | 3.8 | 0.26 | 1200 | K97 | K97 | no | yes |
| ″ | ″ | ″ | ″ | ″ | 0.50 | 850 | ″ | this work | yes | no |
| 46000 | 1.8 | 4.11 | 0.74 | 3.8 | 0.08 | 1200 | K06 | K06 | no | – |
| 44000 | 2.2 | 4.2 | 1.40 | 3.9 | 0.18 | 1200 | P04 | P04 | **yes** | **yes** |
| ″ | ″ | ″ | 0.88 | 3.7 | 0.25 | 360 | this work | this work | yes | no |
| | | | | | NGC 2392 | | | | | |
| 45000 | 2.5 | 4.4 | 0.91 | 3.6 | $\le 0.03$ | 400 | K97 | K97 | no | yes |
| ″ | ″ | ″ | ″ | ″ | 0.32 | 400 | ″ | this work | yes | no |
| 44000 | 2.4 | 4.30 | 0.86 | 3.6 | $\le 0.05$ | 400 | K06 | K06 | no | – |
| 40000 | 1.5 | 3.7 | 0.41 | 3.7 | 0.018 | 420 | P04 | P04 | **yes** | **yes** |

Table 1 lists the stellar parameters for the two test models chosen.[7] Figures 5 and 6 show the profiles of the most important hydrogen and helium lines used in the optical analyses, computed for these two models using WM-basic (solid lines) and Fastwind (dashed lines).

With regard to the implementation of Stark broadening in our code WM-basic, these calculated lines are expected to match their counterparts calculated with Fastwind within errors. The lines cover a wide range of the optical spectra, starting with H$\epsilon$ at 3970 Å and ending with H$\alpha$ at 6563 Å, including not only H lines but also He I and He II lines. As can be seen from the figure, the absorption lines match very well, with only small differences that arise from differences in the density structures. (For the purposes of this comparison the line force has not been calculated consistently but instead adapted in such a way that the computed density structure approximates the density structure used by the Fastwind models.) Figure 3 shows the density profiles for the D30 model. Small deviations occur at $\tau_{\rm Ross} \approx 0.00001$ and between $\tau_{\rm Ross} = 0.0001$ and $\tau_{\rm Ross} = 0.001$. These differences contribute to small changes in the occupation numbers that lead to corresponding small differences in the computed line profiles (the fact that slightly different occupation numbers are responsible for the small differences shown is recognized by the slightly deeper cores of these lines; as the occupation numbers are calculated differently by WM-Basic and Fastwind such small differences are to be expected and can of course not be prevented even if density and temperature were exactly the same). The behavior is similar for the D45 model for which the density structures are shown in Figure 4. Again some minor differences in the density structures are apparent between optical depths of $\tau_{\rm Ross} = 0.0001$ and $\tau_{\rm Ross} = 0.001$, leading to small deviations in the line profiles which are somewhat pronounced at an optical depth of $\tau_{\rm Ross} > 0.1$.

[7] Stark broadening is a form of pressure broadening, and is thus only important for lines which are formed in regions of high atmospheric density, such as in the photospheres of dwarfs. In the regions where the lines go into wind emission, the density is too low for Stark broadening to show a significant effect. Thus, in a supergiant atmosphere where the absorption lines become contaminated with emission from the outer regions, a statement about the correct implementation of Stark broadening is not possible, since one would be comparing the implementation of all of the physics operating in those regions. We have therefore compared our Stark profiles for dwarf atmospheres, where in the regions in which Stark broadening becomes important the density structures of WM-basic and Fastwind are in agreement.

## 4. CSPN observational material

The observed UV spectra of the two CSPNs discussed here, NGC 6826 and NGC 2392, are the same as those used by Pauldrach et al. (2004) in their UV analysis. These spectra were obtained from the INES Archive Data Server on the Web now at `http://sdc.laeff.inta.es/ines/`, providing access to IUE Final Archive data.

The optical material was collected from observations using a variety of telescopes and spectrographs: ESO 3.6 m + CASPEC, ESO NTT + EMMI, Isaac Newton 2.5 m (La Palma) + IDS, Palomar echelle, McDonald 2.1 m + Sandiford echelle. These spectrograms were kindly provided by R.-P. Kudritzki (priv. comm.).

## 5. Consistent optical and UV analysis of the CSPNs NGC 6826 and NGC 2392

In this paper we present a comparison of computed and observed optical and UV spectra for two CSPNs. We calculate optical and UV spectra using our improved WM-basic code, and we want to see if (for each CSPN) a set of stellar parameters exists for which both the predicted optical and the predicted UV spectrum simultaneously match the corresponding observations. Basis for the comparison is the analysis of the UV spectra by Pauldrach et al. (2004) and the optical analysis by Kudritzki et al. (1997), who found significantly different stellar parameters (see Table 2). We have chosen the two CSPNs NGC 6826 and NGC 2392 because they represent extreme examples compared to the other objects in the sample investigated by Pauldrach et al. (2004), according to which NGC 6826 has an extremely high mass of 1.4 $M_\odot$ (near the Chandrasekhar limit), whereas NGC 2392 has a mass of only 0.41 $M_\odot$, the smallest mass in the whole sample.

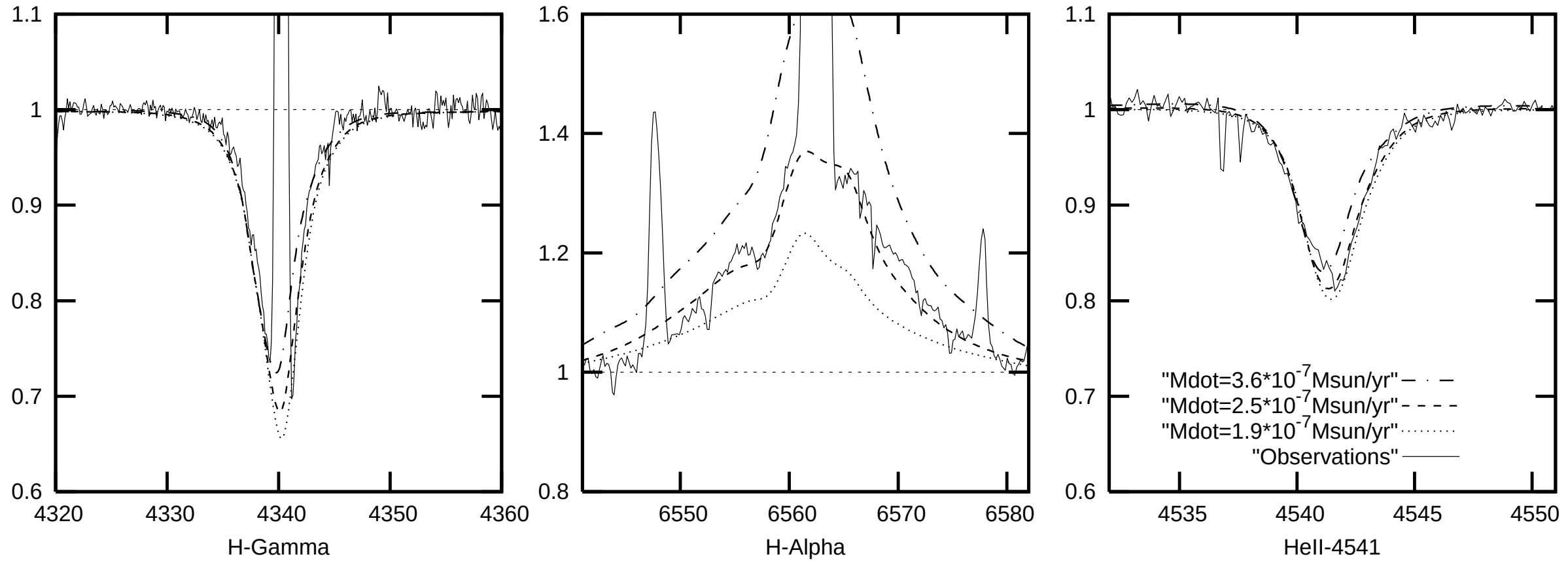

**Fig. 7.** Profiles for the $H\gamma$, $H\alpha$, and He II $\lambda 4541$ lines of the CSPN NGC 6826 calculated with the model atmosphere code Fastwind (J. Puls, priv. comm.) using the stellar parameters derived by Kudritzki et al. (1997), compared to the observed profiles. Note in particular the sensitive dependence of the emission line $H\alpha$ on the mass loss rate. The different predicted line profiles correspond to mass loss rates of $\dot{M} = 1.9{\times}10^{-7} M_{\odot}/\mathrm{yr}$ (dotted), $\dot{M} = 2.5 \times 10^{-7} M_{\odot}/\mathrm{yr}$ (dashed), and $\dot{M} = 3.6 \times 10^{-7} M_{\odot}/\mathrm{yr}$ (dash-dotted).

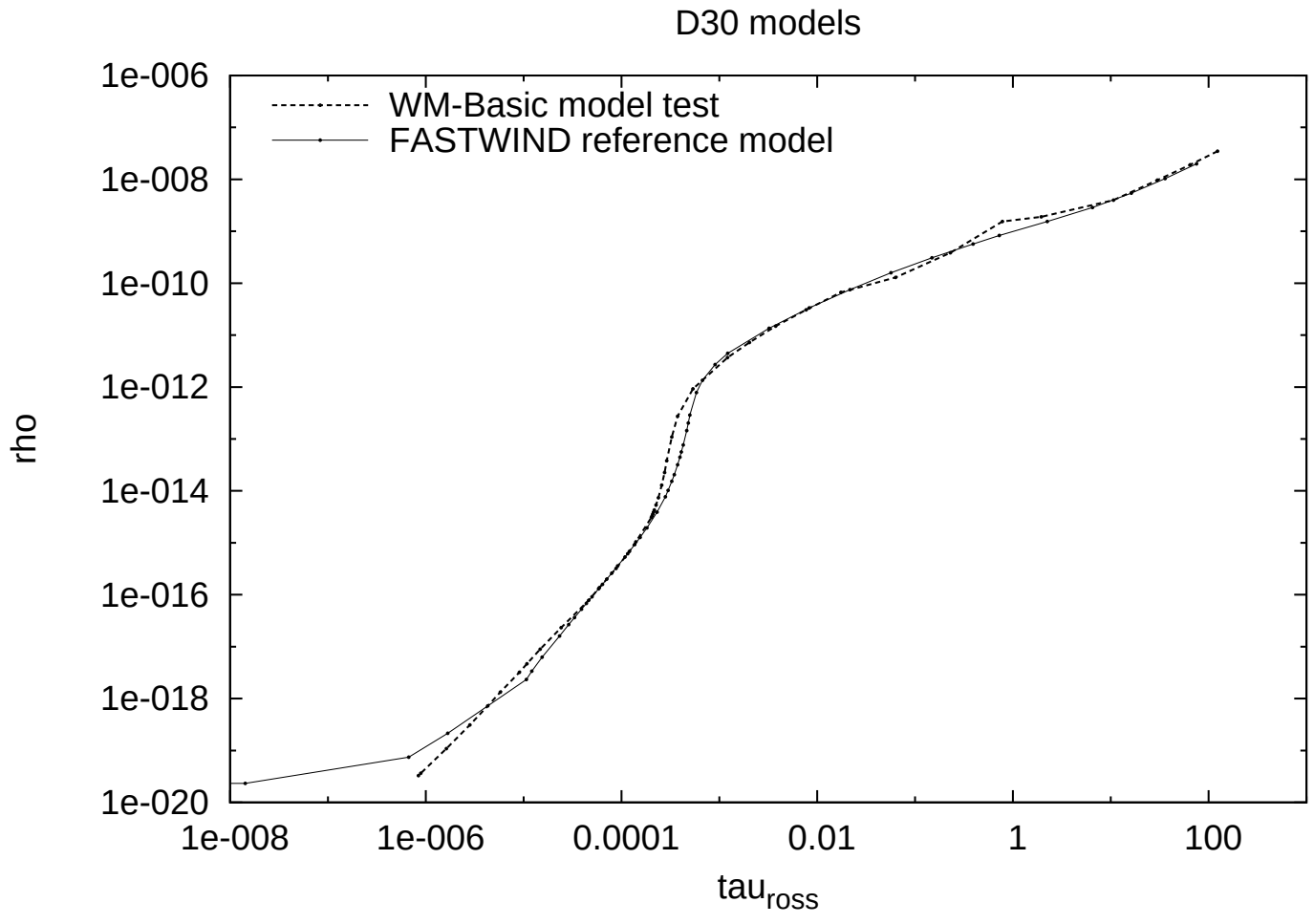

**Fig. 3.** Density profiles for the D30 test model. The solid line shows the density structure used by the Fastwind model, the dashed line shows the hydrodynamically calculated density structure of the WM-basic model, using a (not necessarily consistent) parametrization of the radiative force designed to approximate the density structure of the Fastwind comparison model.

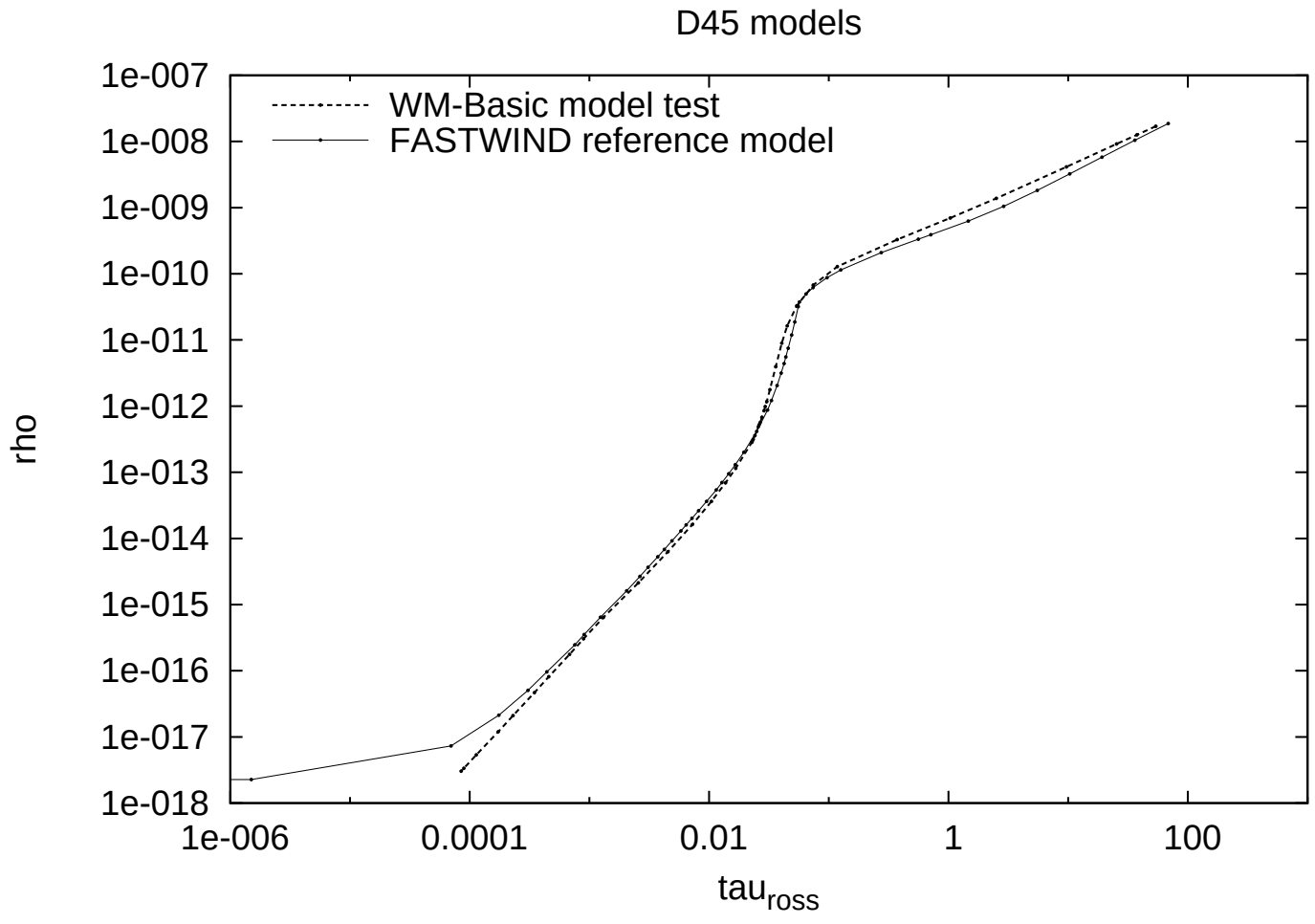

**Fig. 4.** Density structure for the D45 test model. As for the D30 model, the D45 models show small differences in their density profiles compared to one another, leading to some expected differences in the line profiles.

### 5.1. NGC 6826

Kudritzki et al. (1997) have shown their computed $H\gamma$, $H\alpha$, and He II $\lambda 4541$ line profiles for NGC 6826. Their model was calculated with a previous version of the Fastwind model atmosphere code, but calculations (J. Puls, priv. comm.) using the current Fastwind code yield almost identical line profiles (Figure 7). The Kudritzki et al. (1997) determination of the mass loss rate was based on modelling of the $H\alpha$ line. As shown in Figure 7, a mass loss rate of around $\dot{M} = 2.5 \times 10^{-7}\ M_{\odot}/\mathrm{yr}$ yields the best fit to the observed $H\alpha$ line profile for the stellar parameters assumed by Kudritzki et al. (1997) (which were taken to conform to the predicted mass–luminosity relation of theoretical post-AGB evolutionary models).

Figure 8 shows the optical lines from the corresponding WM-basic model run. For this we used the same stellar parameters, but we had to artificially adjust the line force to obtain the desired mass-loss rate of $\dot{M} = 2.6 \times 10^{-7}\ M_{\odot}/\mathrm{yr}$. With this mass loss rate, our calculated absorption lines are almost identical to those presented by Kudritzki et al. (1997). The small deviations in the emission lines $H\alpha$ and He II $\lambda 4686$ are, as already discussed above, due to the differences in the density structure. These arise from the fit parameter $\beta$ (see Section 2.2) used in Fastwind, which effectively allows varying the radius–density relationship in the wind, and thus allows the user to tune the amount of emission in these lines. (Remember that $H\alpha$ and He II $\lambda 4686$ are recombination lines and as such are strongly density-dependent.) Our WM-basic code, on the other hand, does not provide such a free parameter, since the radial run of density and velocity is not chosen by the user but computed from the hydrodynamic equations and the radiative force.[8]

[8] From our point of view it is inherently dangerous to fit line profiles by simply tuning a parameter that describes an assumed velocity law, because it involves the risk of covering up intrinsic weaknesses of other parts of the model description. Thus, being able to reproduce observed line profiles by adapting such a parameter tells us nothing whatsoever about the real physics of the expanding atmosphere, and only serves to fool oneself into believing one has actually made physical progress. It should be realized that a velocity law doesn't have to be assumed for O star atmospheres, because it can be calculated from well-known physical laws. Already in 1975, Castor, Abbott, & Klein have shown the way to proceed, obtaining already reliable results as demonstrated by Pauldrach et al. (1986). An arbitrarily chosen fudge factor to describe

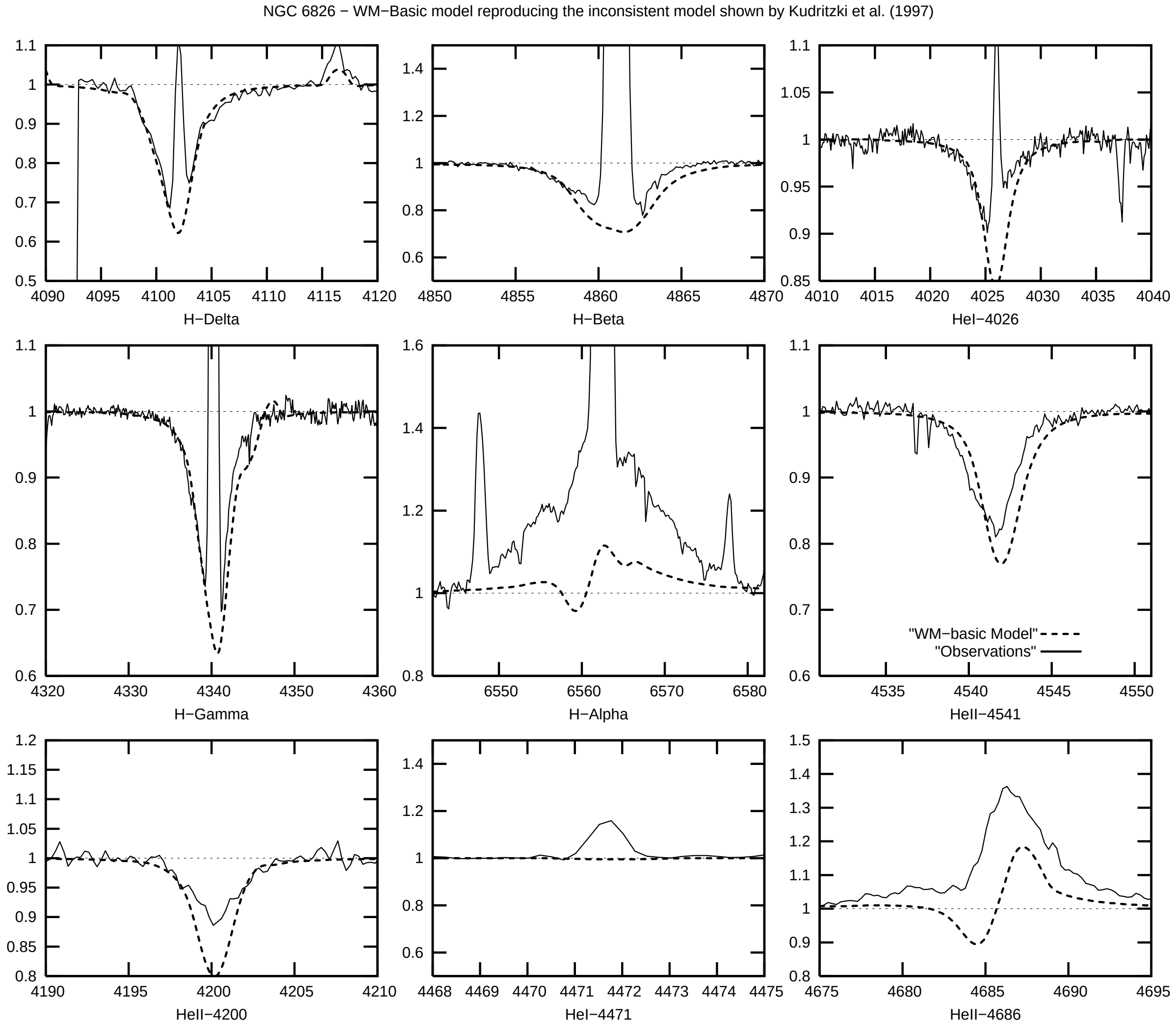

**Fig. 8.** Optical line profiles from our WM-basic model of NGC 6826 with artificially adapted line force reproducing the NGC 6826 model by Kudritzki et al. (1997). Comparing our line profiles to those shown by Kudritzki et al. (their Figure 1, cf. also our Figure 7) we find H$\gamma$ and He II $\lambda$4541 to be almost identical, whereas our H$\alpha$ line is somewhat weaker. All other absorption lines match the observations in good agreement.

Compared to the parameters obtained by Kudritzki et al. (1997), the UV analysis by Pauldrach et al. (2004) had yielded almost the same value for the mass loss rate ($\dot{M} = 1.8 \times 10^{-7}\ M_\odot$/yr), but along with a consistently determined radius and mass. The latter point being the key factor in the analysis of Pauldrach et al. (2004), one of our main interests in the current work was a comparison of two *consistent* models using the parameters sets of Pauldrach et al. (2004) and Kudritzki et al. (1997). The former model is of course identical to that presented by Pauldrach et al. (2004), but here we can now also compare the predicted optical line profiles (see below). For the latter model we have iterated hydrodynamics together with NLTE and line-force calculations to consistency, obtaining a (much larger) mass loss rate of $\dot{M} = 5.0\times10^{-7}\ M_\odot$/yr and a (much smaller) terminal velocity of $v_\infty = 850$ km/s.

Figure 9 shows the predicted line profiles for the selfconsistent model using the Kudritzki et al. (1997) stellar parameters. With the exception of He I $\lambda$4026, the comparison to the observed line profiles of NGC 6826 is unsatisfactory: wind emission begins to fill up the absorption lines, and the emission in H$\alpha$ and He II $\lambda$4686 is much too strong. The bottom part of Figure 9 additionally shows the predicted UV spectrum, compared to the observed UV spectrum. What was already evident from the optical line profiles, namely that the stellar parameters of this model yield a much too large mass loss rate to reproduce the observations, is confirmed by the synthetic UV spectrum: the (unsaturated) P-Cygni lines He II $\lambda$1640 and N IV $\lambda$1719 are much too strong, as is the entire "forest" of Fe and Ni lines spanning the range from 1250 Å to 1500 Å.

We now turn to the model using the Pauldrach et al. (2004) parameters. As described in Section 2.1, these parameters were determined by varying mass and radius (and effective temperature) until the predicted UV spectrum from a *consistent* model with those parameters matched the observed UV spectrum (Figure 10, bottom panel). But with regard to the central question concerning the true stellar parameters of this object, the significant result of our current investigation is that the predicted optical line profiles *also* match the observed optical spectrum *as well as* those of the model using the artificial wind parameters and the Kudritzki et al. (1997) stellar parameters, whereas the *consistent* model using the Kudritzki et al. (1997) stellar parameters utterly fails to reproduce both UV and optical observations.

the velocity structure has therefore not been necessary for more than 35 years.

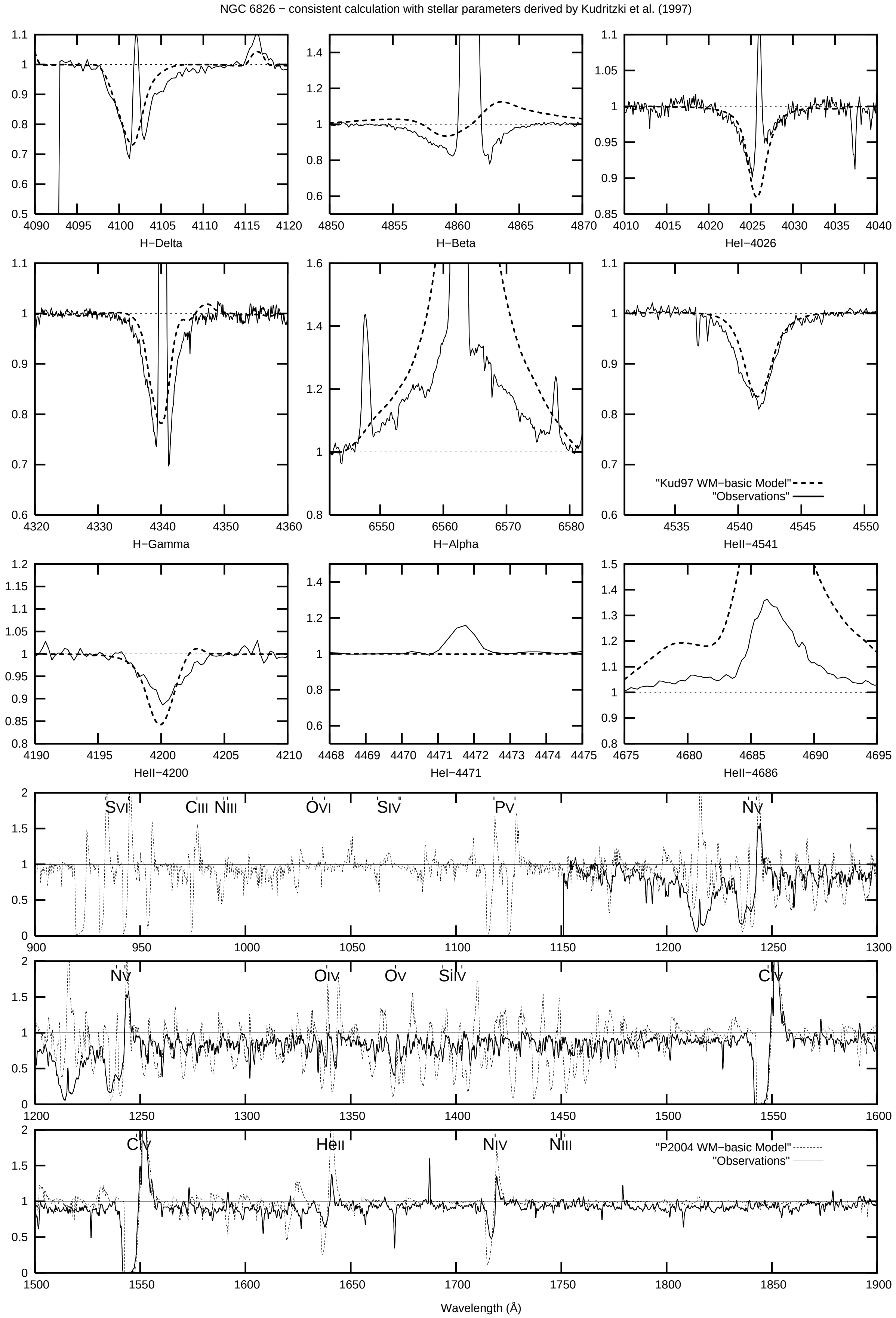

**Fig. 9.** Optical and UV spectra for NGC 6826 from a model based on the stellar parameters from Kudritzki et al. (1997), but with wind parameters consistent to the stellar parameters. Compared to the observations (solid line), the model (dashed) yields far too strong emission lines in both the optical and the UV spectral ranges.

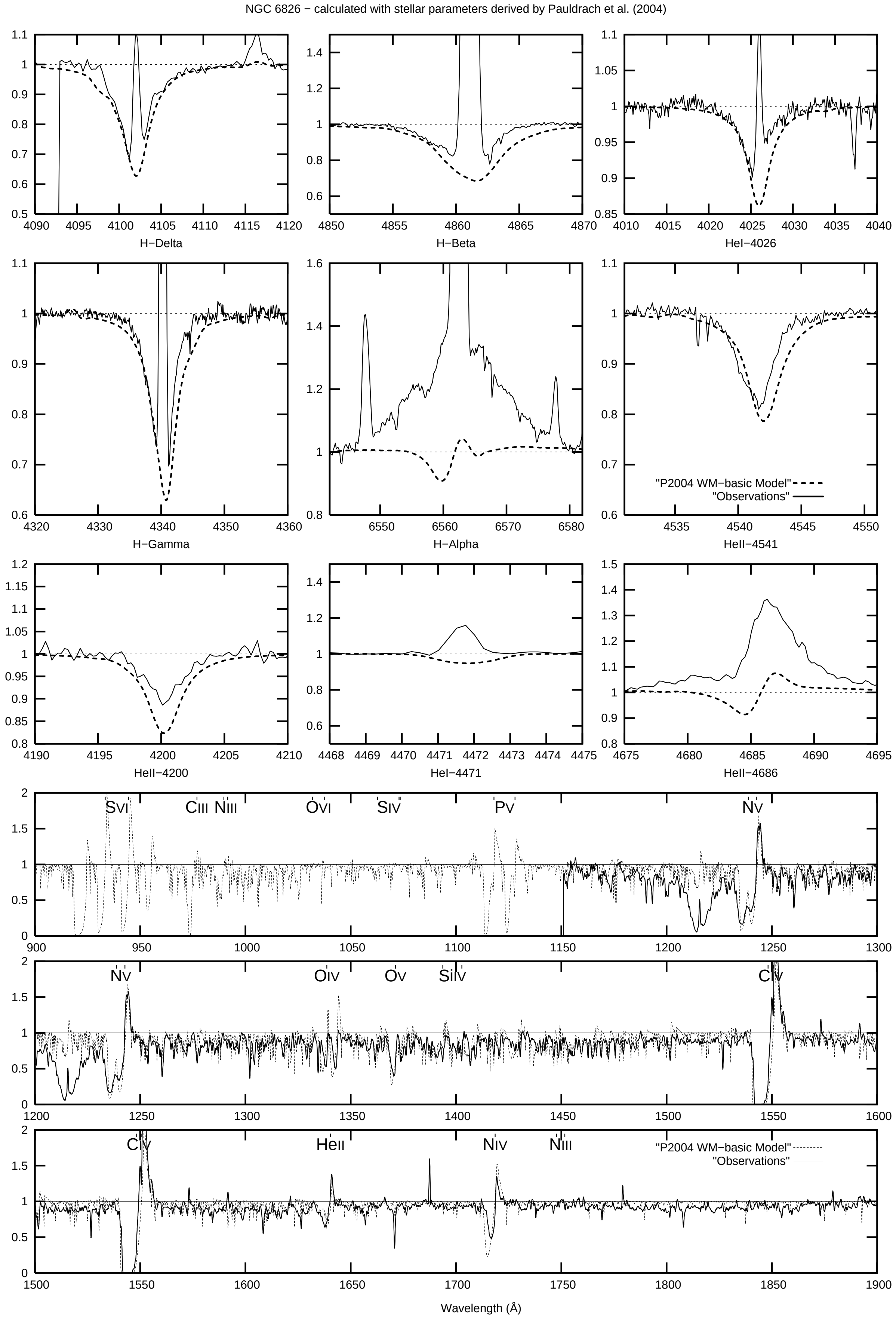

**Fig. 10.** Optical and UV spectra for NGC 6826 from a model using the parameters from Pauldrach et al. (2004) (i.e., stellar parameters from Pauldrach et al. and wind parameters consistent with those stellar parameters). The predicted spectra match the observations as well as those of the model using the artificially adapted wind parameters and the stellar parameters of Kudritzki et al. (1997), but the model shown here has the merit of being hydrodynamically consistent.

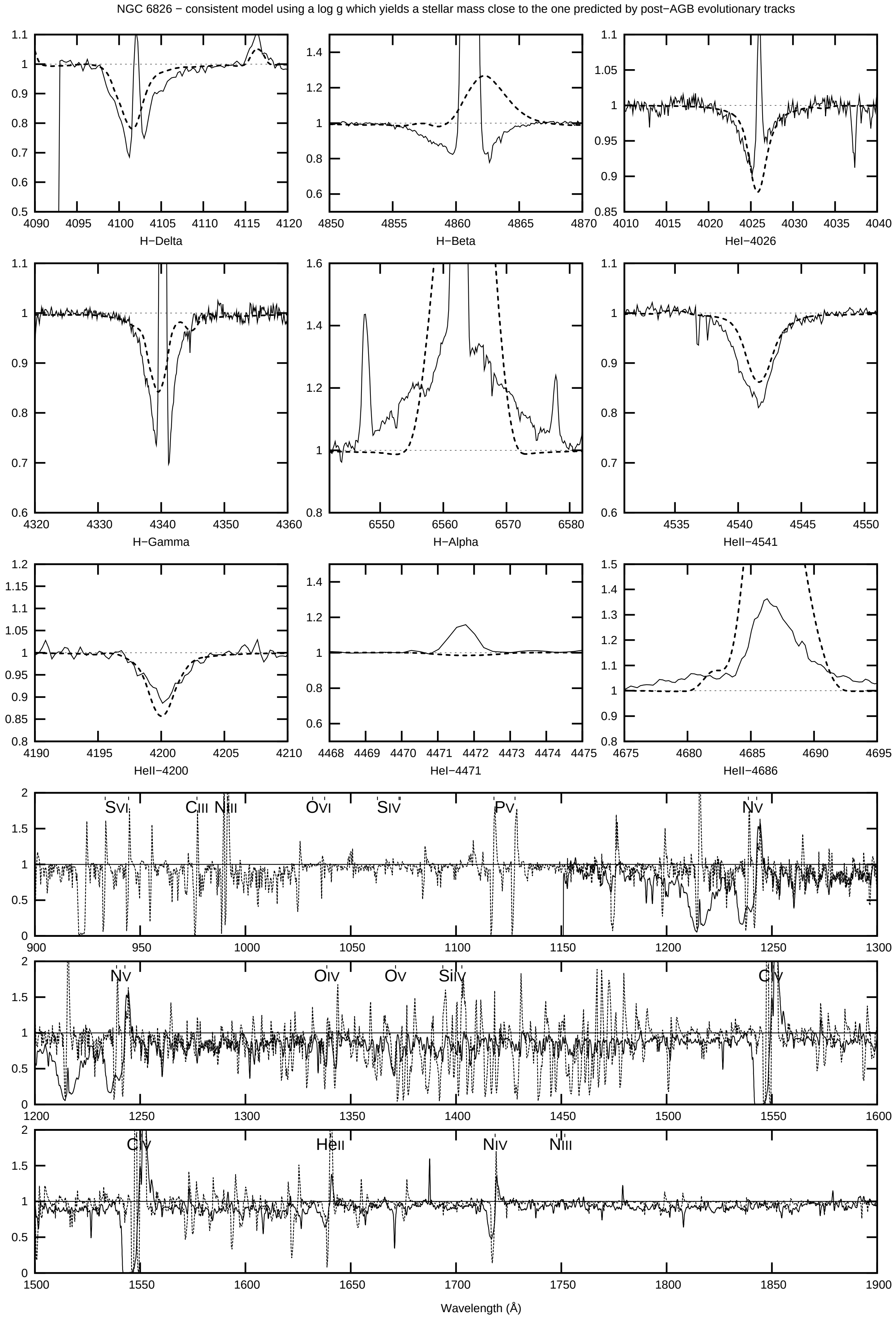

**Fig. 11.** As Figure 10, but using a log $g$ of 3.7, representing a consistent model with a mass very close to the one given by Kudritzki et al. (1997) (see Table 2). The now far too strong emission in He II $\lambda$4686 and H$\alpha$ obviously does not match the observations.

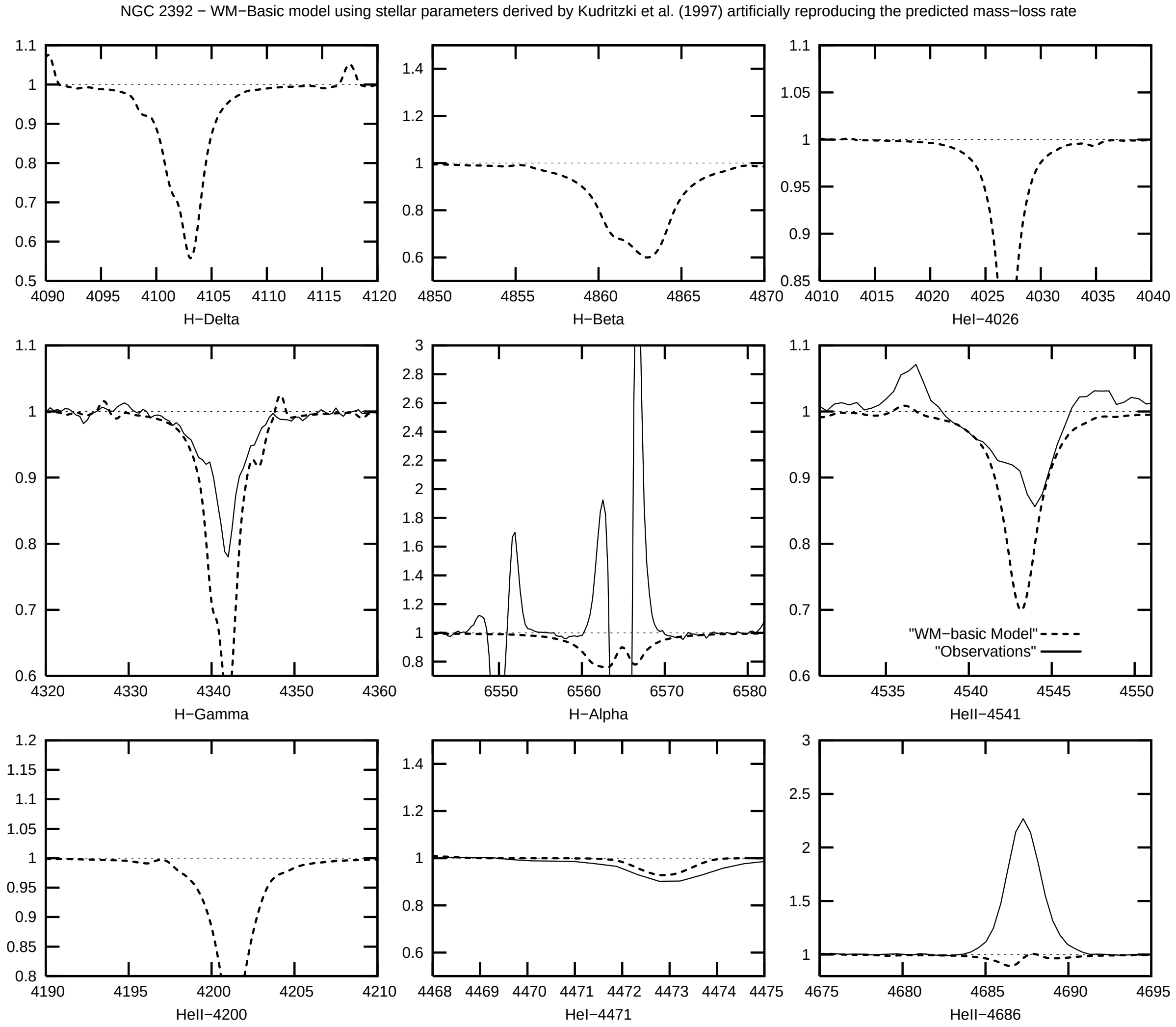

**Fig. 12.** WM-basic model of NGC 2392 with artificially adapted line force reproducing the NGC 2392 model by Kudritzki et al. (1997). Given that our model for NGC 6826 adequately reproduced the line profiles shown by Kudritzki et al. (1997) for that object, we will equivalently assume the same for this CSPN. (Kudritzki et al. (1997) did not show any line profiles for NGC 2392.)

The conclusion to be drawn from this fact is that a match of predicted and observed spectra from any particular model does not immediately guarantee the correctness of the stellar parameters used in that model, since it is very likely that for any set of intermediate stellar parameters an (inconsistent) mass loss rate may be found that yields a similarly adequate fit to the observed spectrum. Of this series of models the singularly distinguished model is the one in which the wind parameters are consistent with the stellar parameters, and the parameters of this model must be regarded as being closest to the true parameters.

The calculated H$\gamma$ profile shown in Figure 10 has wings slightly broader than the observed profile, and it might thus appear that the gravity used in the model is higher than needed. In fact, the careful reader may have noticed that a drop in $\log g$ of only 0.2 (with all other parameters fixed) will bring the mass of this model very close to the mass of the Kudritzki et al. (1997) model for NGC 6826 (see Table 2). While the wings of H$\gamma$ from a consistent model with these stellar parameters now indeed match the observation better (see Figure 11), all other features in both the optical and the UV spectrum are completely ruined. The computed mass loss rate of $\dot{M} = 0.25 \times 10^{-6}\ M_{\odot}/\mathrm{yr}$ incidentally comes close to that used by Kudritzki et al. (1997), but the corresponding terminal velocity of $v_\infty = 360$ km/s is far too small (which can be seen in the narrow line profiles in the UV as well as the narrow optical emission lines), and the resulting higher wind density leads to far too strong emission in H$\alpha$, H$\beta$, and He II $\lambda$4686, as well as filling up the optical absorption lines.

Thus, although the reduction of $\log g$ achieved the goal of a better fit of the H$\gamma$ wings, the model became completely unacceptable with respect to all other spectral features. We must conclude from this that the wings of H$\gamma$ cannot by themselves be considered a reliable indicator of the surface gravity. This conclusion is underscored by a comparison of the H$\gamma$ line profiles shown in Figures 8 and 9, which result from models that have identical stellar parameters, in particular also identical $\log g$. The only difference between the two models is the density structure. Thus, while it is true that the wings of H$\gamma$ are an indicator of the density, the density structure is not determined by the surface gravity alone, but also by the radiative forces, and therefore it is inherently dangerous to base the determination of the surface gravity on a single line which is known to be also influenced by the back-reaction of thousands of other lines.

Nevertheless, the optical *emission* lines of the consistent model using the Pauldrach et al. (2004) parameters do not match the observations (though those of our model using the Kudritzki

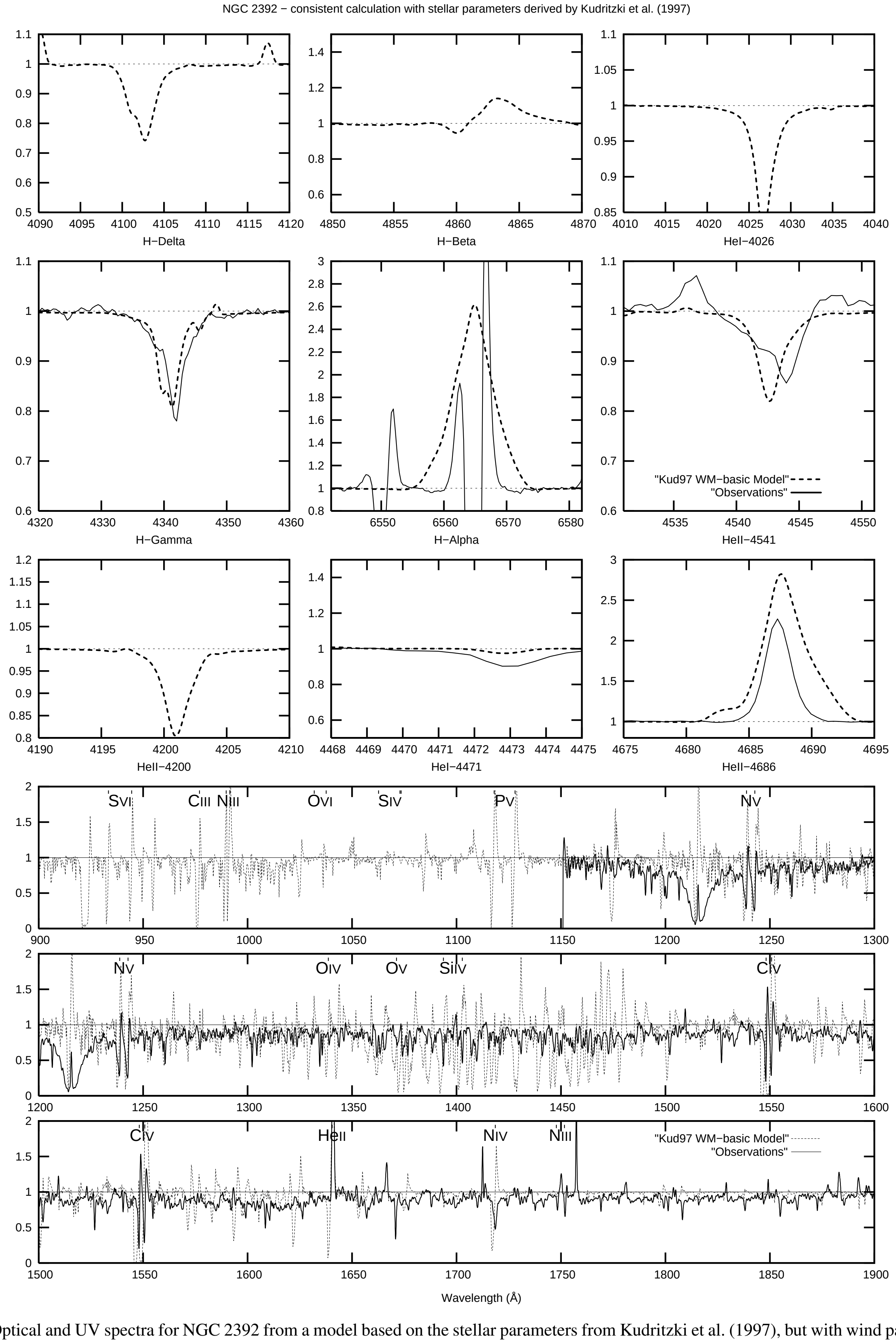

**Fig. 13.** Optical and UV spectra for NGC 2392 from a model based on the stellar parameters from Kudritzki et al. (1997), but with wind parameters consistent to the stellar parameters. Compared to the observations, the model yields far too strong emission lines in both the optical and the UV spectral ranges.

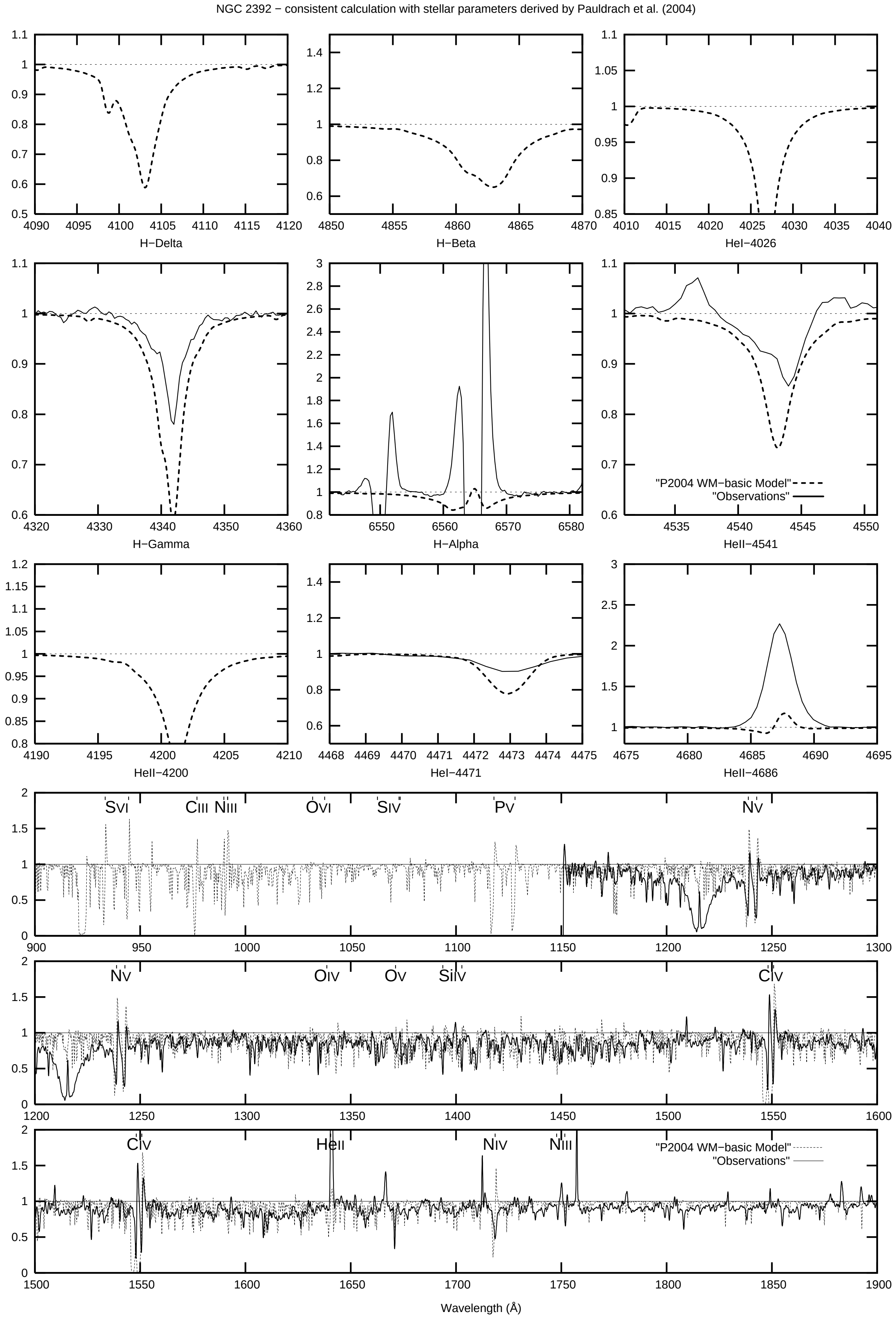

**Fig. 14.** Optical and UV spectra for NGC 2392 from a model using the parameters from Pauldrach et al. (2004) (i.e., stellar parameters from Pauldrach et al. and wind parameters consistent with those stellar parameters). The predicted UV spectrum matches the observations, and the predicted optical line profiles are almost identical to those from the Kudritzki et al. (1997) model.

et al. (1997) stellar and wind parameters do not, either), and this detail merits discussion. Test calculations have shown that the mass loss rate needed to bring H$\alpha$ into agreement is roughly only a factor of two larger than that of the model shown, but we cannot choose a model with a larger mass loss rate (with appropriate changes to the stellar parameters to maintain consistency) since this destroys the match of the UV spectrum. (As discussed above, Kudritzki et al. (1997) had the possibility of adjusting the $\beta$ parameter of the velocity field used in their analysis, but we do not consider this a realistic option since the computed line force does not result in a velocity field with the assumed shape used in those calculations.) Does this mismatch of the optical emission lines invalidate the parameters obtained from the consistent UV analysis? We are convinced that it does not, for the following reason:

Earlier analyses of the mass loss rates of massive O stars based on modelling the optical emission lines had seemed to indicate that the mass loss rates of supergiants were larger than those of dwarfs with similar luminosities (Puls et al. 1996), but this assessment has since been revised by the realization that the winds may be "clumped", the increased density in the clumps leading to a stronger emission of the hydrogen and helium recombination lines than in a "smooth" wind with the same mass loss rate. This revision has resulted in bringing the derived parameters into closer agreement with those predicted by radiation-driven wind theory (Puls et al. 2006), and it is likely that the winds of O-type CSPNs are similarly affected by clumping (Urbaneja et al. 2008). As we still have no consistent theory that will quantitatively predict the run and magnitude of this clumping, however, we are at the moment inclined to not rely too heavily on emission lines which are expected to be thus affected.

### 5.2. NGC 2392

The other CSPN we discuss here is NGC 2392. What makes this object so interesting is that the situation for this star is reversed compared to NGC 6826: Pauldrach et al. (2004) had to significantly decrease the luminosity to obtain a match of predicted and observed UV spectrum. They thereby derived a very small mass of only 0.41 $M_{\odot}$, making this the least massive CSPN of their sample.

In Figure 12 we show our computed line profiles for a model with the stellar parameters of Kudritzki et al. (1997) and an artificially adapted line force resulting in the same mass loss rate as fitted by Kudritzki et al. (1997). Unfortunately, the observational material available for this object is rather poor and for some of the lines no observations are available at all. Nevertheless we show our computed line profiles as reference and, since Kudritzki et al. (1997) did not show any line profiles at all for this object but our NGC 6826 model adequately reproduced theirs, we will assume that the profiles computed by Kudritzki et al. (1997) for NGC 2392 were similar to these.

In Figure 13 we show predicted optical line profiles and UV spectrum for a consistent model using the Kudritzki et al. (1997) stellar parameters for NGC 2392. It is evident that the consistently computed mass loss rate of $3.2 \times 10^{-7}$ $M_{\odot}$/yr is much too large to reproduce either UV or optical spectrum. This is in contrast to the consistent model from Pauldrach et al. (2004), shown in Figure 14, which not only agrees much better with the observed UV spectrum, but also predicts optical line profiles which are nearly identical to those from the (inconsistent) Kudritzki et al. (1997) model shown in Figure 12.

In summary, the situation for NGC 2392 is similar to that for NGC 6826: the Kudritzki et al. (1997) model with the artificial mass loss rate offers no observational advantage over the Pauldrach et al. (2004) model, and the latter again has the merit of being hydrodynamically consistent. The consistent model using the Kudritzki et al. (1997) stellar parameters, on the other hand, again fails to reproduce both UV and optical observations.

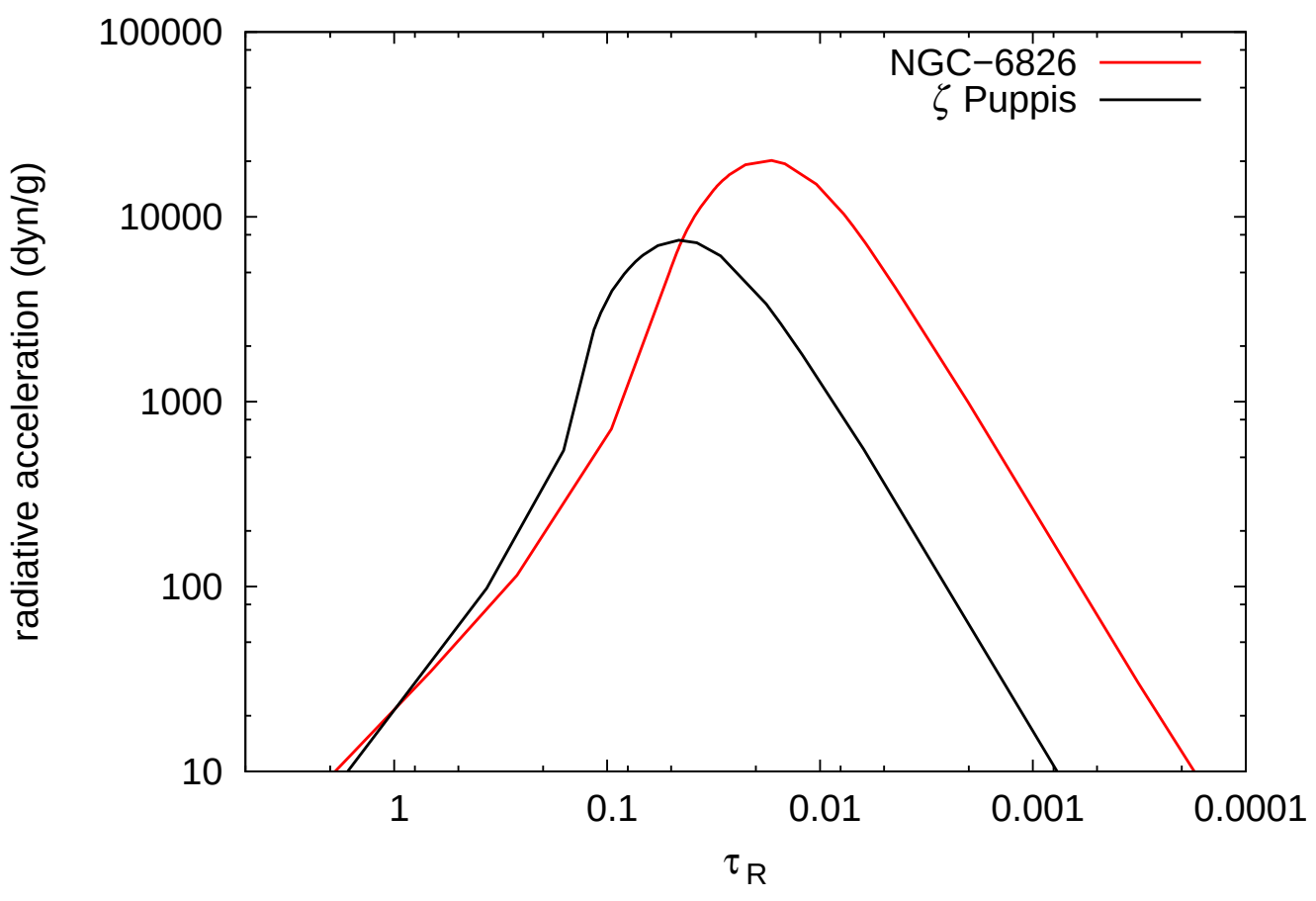

**Fig. 15.** Radiative acceleration versus optical depth in the wind of the CSPN NGC 6826 (shown in red) using the parameters based on the UV analysis (model P04 in Table 2), compared to the radiative acceleration in a massive O star wind ($\zeta$ Puppis; shown in black).

### 5.3. Comparison of the dynamical parameters of CSPN and O star winds

The terminal velocity $v_\infty$ and the mass loss rate $\dot{M}$ are the primary dynamical parameters of a stellar wind and the essential parameters for a consistent theoretical description. Both parameters are connected via the equation of continuity

$$\dot{M} = 4\pi r^2 \rho(r) v(r) \qquad (1)$$

(where the outwards monotonically increasing velocity field $v(r)$ reaches its maximum value at roughly a hundred stellar radii and thus defines the terminal velocity $v_\infty$ of the wind). Both the velocity $v(r)$ and the density $\rho(r)$ are functions of the radial coordinate $r$. Via the competing forces of line-driving and gravity they become unique functions of the basic stellar parameters $R_*$, $L_*$, and $M_*$.

Because our sample of CSPN stars shows pronounced wind features it is a natural step to examine these dynamical parameters and their relation to the basic stellar parameters as an additional and independent point of our investigation. This discussion is thus an extension of the investigation of Pauldrach et al. (1988) who already showed that the calculated terminal wind velocities are in agreement with the observations and therefore allow an independent determination of stellar masses and radii.[9] With respect to this result, different sets of stellar masses and radii applied to our sample of stars should therefore lead at least partly to an inconsistent behavior with regard to the predictions of radiation-driven wind theory. Such an inconsistent behavior can, for example, be evident in the ratio $v_\infty/v_{\rm esc}$ of the terminal wind velocity and the escape velocity of individual stars.[10] In this investigation it is significant to not only compare the $v_\infty/v_{\rm esc}$ ratios obtained with our improved models for the CSPN sample to the corresponding "observed ratios" (which are based on the mass–luminosity relation of CSPNs), but also to the ratios of a "normal" O star sample and its corresponding observations. Because the UV-spectra of the CSPNs of our sample are very similar to those of massive O stars, the application of the theory of radiation-driven winds is expected to also yield similar results for the terminal velocity $v_\infty$ of these CSPNs and the massive O stars. One would therefore expect the line force in the winds of our CSPNs to be of a strength comparable to that of the winds of massive O stars. In view of this expected result we have compared the computed line force from our consistent model of NGC 6826 to the corresponding one of the well known massive O star $\zeta$ Puppis, which has a similar UV spectrum. We show the radial run of the radiative forces, which include all relevant continuum and line contributions, in Fig. 15.

[9] We note that Pauldrach et al. (1988) were also able to show that the winds of CSPNs are driven by radiation pressure and thus the insights of radiation-driven wind theory are applicable here.

Surprisingly, the two curves of the radiative acceleration differ considerably in strength and shape, although the same sophisticated radiation-driven wind modelling has been applied in each case, and the spectra resulting from the models each match the corresponding observed spectra. The key to solving this apparent puzzle lies in understanding the basic relationships between the radiative acceleration and the density and velocity of the outflow.

Figure 16 shows the line acceleration along with the corresponding wind densities $\rho$ as function of the scaled radius $r' = r/R_*$ (the $x$-scale chosen in Fig. 16 serves to emphasize the relevant radial range). In the inner (photospheric) regions the densities are similar, and this similarity in the models is supported by the fact that the optical (photospheric) spectra are also very similar (see Sect. 5.1 and Pauldrach et al. 2012). The similarity of the UV spectra, on the other hand, implies that the relevant wind features are formed at about the same optical depth in the wind. The optical depths in the wind are proportional to

$$\tau_R \propto \frac{\dot{M}}{v_\infty R_*^2} R_* \qquad (2)$$

(cf. Puls & Pauldrach 1991), where

$$\frac{\dot{M}}{v_\infty R_*^2} = \bar{\rho} \qquad (3)$$

may be unterstood as a characteristic wind density (Pauldrach et al. 1990). Similar optical depths in the winds of both stars therefore implies

$$\bar{\rho}^{(\zeta\ \rm Pup)} R_*^{(\zeta\ \rm Pup)} \approx \bar{\rho}^{(\rm NGC\ 6826)} R_*^{(\rm NGC\ 6826)} \qquad (4)$$

and thus

$$\frac{\bar{\rho}^{(\rm NGC\ 6826)}}{\bar{\rho}^{(\zeta\ \rm Pup)}} \approx \frac{R_*^{(\zeta\ \rm Pup)}}{R_*^{(\rm NGC\ 6826)}}. \qquad (5)$$

Since

$$R_*^{(\rm NGC\ 6826)} < R_*^{(\zeta\ \rm Pup)} \qquad (6)$$

this means that

$$\bar{\rho}^{(\rm NGC\ 6826)} > \bar{\rho}^{(\zeta\ \rm Pup)}. \qquad (7)$$

This behavior is indeed also reproduced by the models (see Fig. 16).

[10] Investigations and interpretations of these ratios have already been performed by Pauldrach et al. 1988 (regarding CSPNs) and Pauldrach et al. 1990 (regarding massive O-stars).

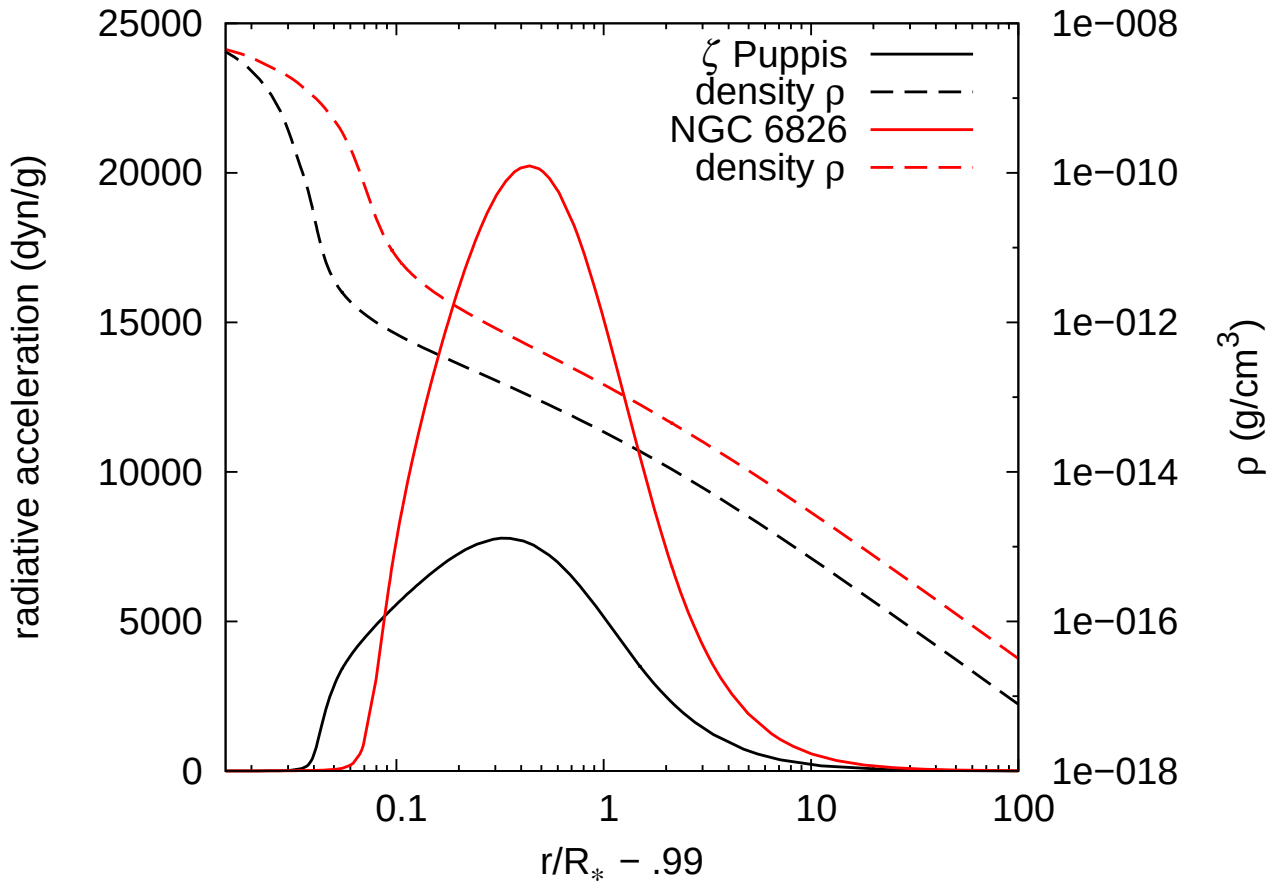

**Fig. 16.** A more detailed illustration of the radiative acceleration for the CSPN NGC 6826 in comparison to the massive O star $\zeta$ Puppis. Additionally the density of both models has been plotted as well. The difference in radiative acceleration of both models shown in Fig. 15 can be well understood by the means of this plot (see text).

As Pauldrach et al. (1990) have shown, a higher wind density leads to a higher radiative acceleration, which explains the behavior shown in Fig. 16. A higher radiative acceleration in turn leads to a higher mass loss rate per unit surface element, i.e., $\dot{M}/R_*^2$ (see Pauldrach et al. 1990), which in turn leads to a smaller terminal velocity (for comparable effective temperatures, cf. Pauldrach et al. (1988), their Figs. 10 and 6a).

The lower observed terminal velocities of CSPNs with pronounced winds thus provide evidence for the higher wind densities of these CSPNs compared to those of massive O stars (see Pauldrach et al. (1988), Pauldrach et al. (2004) and Table 2 of this work). The higher values of the line force for NGC 6826 compared to $\zeta$ Puppis thus follow directly from its smaller radius and the resulting consequences.

**The $v_\infty/v_{\rm esc}$ ratios of CSPNs compared to massive O stars.** The observations show a dependence of $v_\infty$ on $T_{\rm eff}$, which was confirmed by the theory of radiation-driven winds. Additionally, the theory predicts a correlation of $v_\infty$ to the other two stellar parameters, mass $M_*$ and radius $R_*$, via the escape velocity $v_{\rm esc}$ (see Pauldrach et al. 1988),

$$v_{\rm esc} = \left(2GM_*(1-\Gamma)/R_*\right)^{1/2} \qquad (8)$$

(where $\Gamma = L_*/(4\pi cGM_*/(\chi_{\rm Th}/\rho))$ is the ratio of stellar luminosity to Eddington luminosity, $G$ is the gravitational constant, and $\chi_{\rm Th}$ is the Thomson absorption coefficient).

Figure 17 shows the ratio of $v_\infty/v_{\rm esc}$ for our CSPN sample as a function of different stellar parameters, compared to the corresponding values from a sample of massive O stars (cf. Pauldrach et al. 2011).

As is shown, the correlation of $v_\infty$ to the escape velocity $v_{\rm esc}$ does not represent a strict and simple behavior. This fact was already recognized by Howarth & Prinja (1989) from a purely observational point of view. Moreover, in their comprehensive study of the wind properties of a sample of 205 O stars observed with the IUE satellite, Howarth & Prinja (1989) drew attention to a discrepancy between predictions obtained from scaling laws which were based on theoretical models (Pauldrach et al. 1986) and their interpretation of observations. This discrepancy concerned the observed dependence of the ratio of $v_\infty/v_{\rm esc}$ on the stellar parameters $T_{eff}$, $R_*$, and $M_*$. This dependence was not represented by the models of Pauldrach et al. (1986), which yielded $v_\infty/v_{esc}$ ratios in a range of only 2.0 . . . 3.0, whereas the observations revealed ratios in a range of 2.0 . . . 4.5 in the same range of stellar parameters (cf. Howarth & Prinja 1989, their Fig. 10).

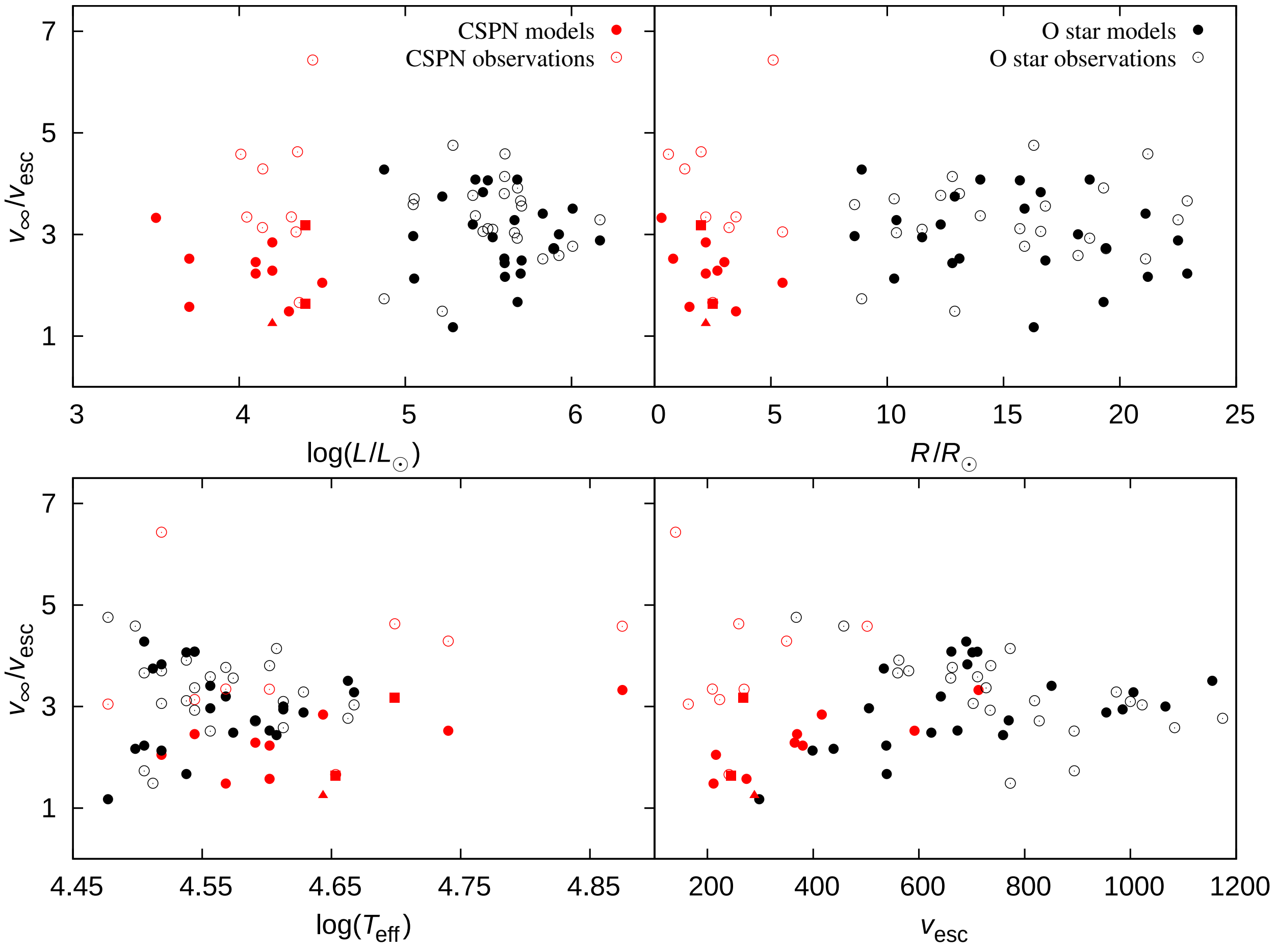

**Fig. 17.** The ratio $v_\infty/v_{esc}$ as a function of the stellar parameters $L_*$, $R_*$, $T_{eff}$, and $v_{esc}$. Compared to the theoretical predictions of Pauldrach et al. (2004) (P04; CSPNs), this work, and Pauldrach et al. (2011) (P11; massive O stars) the values deduced from observations and theoretical relations are shown for our sample of CSPNs (Kudritzki et al. 1997; K97) and a sample of massive O stars (Repolust et al. 2004; R04). The two filled squares shown in the plots depict the two consistent CSPN models for NGC 6826 and NGC 2392 which have been computed on the basis of the parameter sets of K97. The filled triangle represents an additional consistent model for NGC 6826 where the surface gravity has been reduced to a value of $\log g = 3.7$ (the parameters of the models are listed in Table 2). For a discussion see the text.

However, in a revised version of their procedure, Pauldrach et al. (1990) were able to show that for certain parameter ranges of massive O stars the ratio of $v_\infty$ to $v_{esc}$ depends sensitively on the stellar parameters,

$$v_\infty/v_{esc} = f(T_{eff}, Z, R_*, M_*) \quad (9)$$

(where $Z$ denotes the metallicity). This ratio does therefore not present a simple linear function but can vary tremendously from star to star, producing a significant overall scatter (in a range of 2.4 . . . 5.1).[11] Pauldrach et al. (1990) explained this behavior as a result of differences in the ionization structure in the wind, which in particular causes back-reactions on the line force from changes in the level populations via the non-linear behavior of the strong UV line blocking (in particular the back-reaction on and of iron (Fe) has a strong influence in this regard).

[11] We note that based on a less consistent procedure of modelling CSPN winds a similar range of the ratios of $v_\infty$ to $v_{esc}$ was also found by Pauldrach et al. (1988), albeit without the scatter.

A change of the stellar parameters therefore leads to an increase of the mass-loss rate which in turn increases the UV line blocking, which decreases the radiative temperatures in relation to the effective temperature and which thus increases the population of the lower ionization stages. As these lower ionization stages have more strong lines at the corresponding radiative flux maximum (Pauldrach 1987), the net radiative force is increased, and this behavior reinforces the increase of the UV line blocking and leads to a change of the terminal velocity even if the escape velocity $v_{esc}$ has not changed.

Regarding the massive O stars, the results of our current investigation (see Fig. 17) are principally in accordance with those earlier analyses, and, as expected, the improved consistency of our current models does not fundamentally change this behavior, leading now to a scatter of 1.3 . . . 4.4 in $v_\infty/v_{esc}$, in the same range as the observed values.

As another expected result Fig. 17 shows a similar spread in $v_\infty/v_{esc}$ for our CSPN models (the CSPNs are clearly separated from the massive O stars in the upper two panels of Fig. 17 due to the very different radii of these two groups of stars). But not only is the spread roughly the same, also the mean value of the calculated $v_\infty/v_{esc}$ ratios turns out to be not very different for both groups of stars (being somewhat smaller for the CSPN models). Even this is expected from our discussion above of the radiative acceleration which gives rise to considerably lower terminal velocities of CSPNs with pronounced winds because of their higher wind densities as a result of their much smaller radii. Thus, although the CSPNs of our sample have somewhat reduced escape velocities $v_{\rm esc}$ compared to the massive O stars (see below), with their range of $1.2\ldots3.3$ of the $v_\infty/v_{\rm esc}$ ratios the model calculations behave as expected. And this is also the case for the somewhat decreased (compared to the range of O stars) lower limit of 1.2 of $v_\infty/v_{\rm esc}$.

As is shown in the bottom right panel of Fig. 17 the mean value of the escape velocities $v_{\rm esc}$ themselves (not the ratios) of the CSPN models and observations is slightly smaller than the ones of the massive O star observations and models. Obviously, this must be a consequence of the dependence of $v_{\rm esc}$ on the stellar mass $M_*$ and the radius $R_*$ of the objects (cf. Eq. 8). However, $v_{\rm esc}$ also depends on $\Gamma$, the ratio of stellar luminosity to Eddington luminosity. The behavior of this parameter is illustrated in Fig. 18, which shows the calculated mass loss rates $\dot{M}$ versus $\Gamma$ for both the CSPN and the O star sample. The average $\Gamma$ values are clearly smaller for the CSPNs than for the massive O stars. The interpretation of this behavior is straightforward, due to the proportionality of $\Gamma$ to $L_*/M_*$: this ratio becomes smaller for CSPNs than for massive O stars because the difference in $L_*$ is much larger than the difference in $M_*$, and therefore CSPNs are in general farther away from the Eddington limit than massive O stars are. But a decreasing value of $\Gamma$ implies an increasing value of $v_{\rm esc}$, since $v_{\rm esc}$ is proportional to $(1-\Gamma)$, and this behavior may at first appear to be in contradiction with the finding that the escape velocities are on average smaller for the CSPNs than for the massive O stars (bottom right panel of Fig. 17). The apparent contradiction is, however, resolved by recognizing that going from massive O stars to CSPNs the $M_*/R_*$ ratio decreases more strongly than $1/(1-\Gamma)$.

As an important result of Fig. 17 we thus find that, despite the reduction of the escape velocities $v_{\rm esc}$, somewhat smaller $v_\infty/v_{\rm esc}$ ratios are obtained for our CSPNs compared to massive O stars, because the lower terminal velocities $v_\infty$ observed (and calculated) for our CSPNs dominate this effect.[12]

We are now well-prepared to draw the major conclusion of this section from Fig. 17. This conclusion concerns the range of the "observed" CSPN $v_\infty/v_{\rm esc}$ ratios derived from a combination of observations and theoretical relations (e.g., the post-AGB core-mass–luminosity relation, CMLR). As shown in Fig. 17, for the massive O star observations the range of the $v_\infty/v_{\rm esc}$ ratios is in good agreement with the range obtained for the corresponding model calculations. The CSPN model calculations also yield ratios which lie in the range of the $v_\infty/v_{\rm esc}$ spread obtained for the massive O stars (see the discussion above). Clearly at odds, however, are the CSPN "observations", yielding much too high $v_\infty/v_{\rm esc}$ ratios in general, and in particular giving values near 5 or even higher for almost half of the sample! This discrepant behavior is noticeable in all panels of Fig. 17, and appears conspicuously eye-catching in the upper right panel.

This result is especially interesting because such a discrepancy does not emerge when comparing observations and model calculations in regard to the wind-momentum–luminosity relation[13] (WLR), one of the most fundamental relations predicted by the theory of radiation-driven winds. No significant differences between the behavior of the observed wind momenta of massive O stars and CSPNs with pronounced winds are seen, and the computed wind momenta of both the O star models (Pauldrach et al. 2011) and the CSPNs (Pauldrach et al. 2004) agree well with the corresponding observations. Furthermore, the wind momenta of the CSPN models and observations lie along the line extrapolated from the massive O stars (see Pauldrach et al. 2004, their Fig. 6). This result confirms the finding of Pauldrach et al. (1988) and Kudritzki et al. (1997) that the winds of both groups of hot stars are radiation-driven.

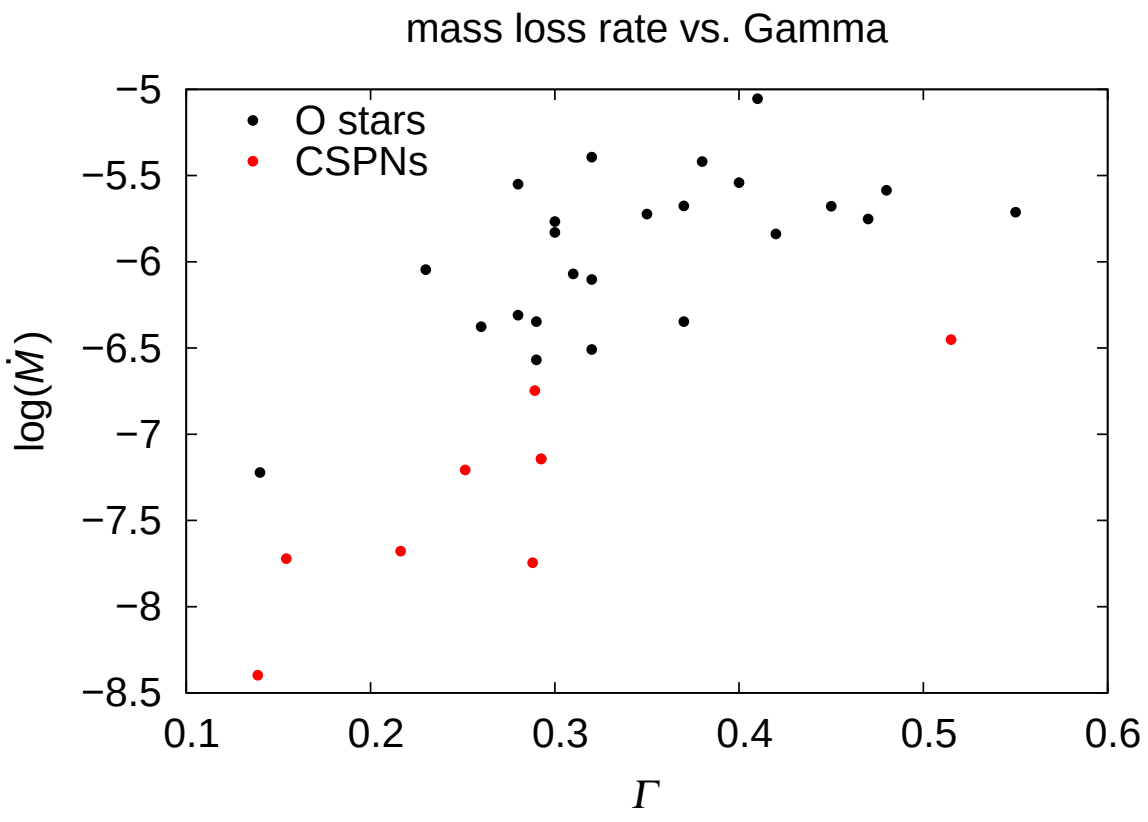

**Fig. 18.** Predicted mass loss rates for our samples of CSPNs and O stars versus $\Gamma$, the ratio of stellar luminosity to Eddington luminosity.

So there is obviously an intrinsic difference between the information provided by the WLR and the information contained in the $v_\infty/v_{\rm esc}$ ratios. This difference regards the mass of the objects. The wind momenta are essentially independent of the stellar masses (a larger $v_\infty$ coincides with a smaller $\dot{M}$ and vice versa; see Kudritzki et al. 1995 and Puls et al. 1996), so any possibly erroneous assumption about the masses made in the modelling of the objects or the interpretation of the observations does not enter into the diagnostic quantity. This is not the case, however, when considering the $v_\infty/v_{\rm esc}$ ratios, which are explicitly mass-dependent. Furthermore, we must realize that the much too high $v_\infty/v_{\rm esc}$ ratios obtained for our CSPN "observations" are not due to the numerators ($v_\infty$), which are directly observable quantities (measurable from the blue edge of the saturated P-Cygni profiles in the UV spectra), and must therefore be due to the denominators ($v_{\rm esc}$). The high ratios are even more significant because from the discussion above we would have expected the highest $v_\infty/v_{\rm esc}$ values to occur for some massive O stars rather than for some CSPNs.

The understanding of this behavior is clearly coupled to the fact that, in contrast to the terminal velocities $v_\infty$, the escape velocities $v_{\rm esc}$ appearing in the "observed" CSPN $v_\infty/v_{\rm esc}$ ratios are not directly observable quantities. The escape velocities are connected to the masses and the radii of the objects, and these quan-tities had (as is usual) been assumed by Kudritzki et al. (1997) to conform to the theoretical CMLR. This is of course not a problem for the WLR, which shows agreement between the predictions of radiation-driven wind theory and the observations since the wind momenta are independent of the mass of the objects.

[12] In the bottom left panel of Fig. 17 we plot the ratio of $v_\infty/v_{\rm esc}$ versus the effective temperature $T_{\rm eff}$. As shown by Pauldrach et al. (1988) and Pauldrach et al. (1990), the theory of radiation-driven winds predicts a mutual dependence between $v_\infty$ and $T_{\rm eff}$. Since the range of their effective temperatures is almost the same, no significant difference between the corresponding results of both stellar groups should therefore be observed. And this is indeed the case: the characteristic values of the $v_\infty/v_{\rm esc}$ ratios and the effective temperatures $T_{\rm eff}$ differ only slightly between our CSPNs and the massive O stars.

[13] The wind-momentum–luminosity relation is a simple relation between the quantity $\dot{M}v_\infty$, which has the dimensions of a momentum loss rate, and the stellar luminosity (Lamers & Leitherer 1993, Kudritzki et al. 1995): the mechanical momentum of the wind flow ($\dot{M}v_\infty$) results from the transfer of momentum from the radiation field to the gas via photon absorption in metal lines (which defines the driving mechanism of the wind) and is thus mostly a function of photon momentum ($L/c$) and therefore related to the luminosity.

With respect to these considerations it is legitimate to conclude that there is a problem with the core-mass–luminosity relation at its high-mass end. We emphasize that we draw this conclusion purely from a comparison of the terminal velocities, independent of the behavior of $\dot{M}$ and the interpretation of either the UV or the optical spectra. We further note that consistent wind model calculations based on stellar parameters conforming to the CMLR also yield $v_\infty/v_{\rm esc}$ ratios which lie in the expected range (however, the terminal velocities obtained by these models are highly incompatible with the observations).[14] We thus conclude that the extraordinarily high values of $v_\infty/v_{\rm esc}$ appear just in cases where the numerator is disconnected from the denominator, i.e., in cases where the observed terminal velocity is not consistent with the assumed mass and radius of the star.

Taking our results of the investigation of the dynamical parameters together with our results obtained from the spectral analysis, and as a third point the assumption of the CMLR for post-AGB stars, we find that only two of these considerations are intrinsically reconcilable, whereas the third consideration is irreconcilable with this pair. For our sample of CSPN objects this means in particular:

- Analyses aiming at reproducing the observed optical and UV spectra based on masses derived from the CMLR yield dynamical wind parameters which are not consistent with the corresponding observations.
- Analyses employing a consistent treatment of the dynamical wind parameters based on masses derived from the CMLR yield optical and UV spectra which do not match the observed spectra.
- Analyses employing a consistent treatment of the dynamical wind parameters aiming at reproducing the observed optical and UV spectra yield masses which are not in accordance with the CMLR.

Moreover, for four of the CSPNs shown in Fig. 17 the $v_\infty/v_{\rm esc}$ ratios that follow from the parameters derived by Kudritzki et al. (1997) based on the CMLR are so extraordinary large that they are not reconcilable with the theory of radiation-driven winds, which on the other hand is strongly supported by a comparison to the WLR.

## 6. Summary and conclusions

In order to obtain further constraints on the true stellar parameters of CSPNs we have applied our stellar atmosphere code WM-basic to two objects from the sample of CSPNs investigated earlier by Kudritzki et al. (1997) and Pauldrach et al. (2004). We have chosen stars for which the two analyses yielded strongly differing parameters, namely NGC 6826 and NGC 2392, for which Pauldrach et al. (2004), comparing observed UV spectra with those predicted by hydrodynamically consistent models, had derived masses of 1.4 $M_\odot$ and 0.41 $M_\odot$, respectively, whereas Kudritzki et al. (1997), working with optical spectra and assuming that the mass–luminosity relation of theoretical post-AGB evolutionary models was valid, had obtained roughly equal masses of 0.92 $M_\odot$ and 0.91 $M_\odot$.

As a prerequisite to this analysis we had extended the applicability of WM-basic to the simultaneous analysis of UV and optical lines by implementing Stark broadening and testing the improved code by comparing the computed optical line profiles with those from two reference O star models calculated with the well-known model atmosphere code Fastwind, showing the results to be in excellent agreement, with only small differences in the predicted line profiles attributed to differences in the density structures of the two codes.

Our results regarding the stellar parameters of CSPNs confirm the conclusion tentatively drawn by Pauldrach et al. (2004), namely that the contradiction between the stellar parameters derived by Kudritzki et al. (1997) and Pauldrach et al. (2004) is not the result of using optical spectra in the former analysis and UV spectra in the latter, but is instead due to the *missing consistency* between stellar and wind parameters in the analysis of Kudritzki et al. (1997).

We arrive at this conclusion because we have now shown that the hydrodynamically consistent models of Pauldrach et al. (2004) reproduce not only the observed UV spectra but also yield optical line profiles which are nearly identical to those from models using the stellar parameters of Kudritzki et al. (1997) and the artificial (i.e., inconsistent) radiative line force necessary to reproduce the wind parameters fitted by Kudritzki et al. (1997). *Consistent* models using the Kudritzki et al. (1997) stellar parameters, on the other hand, reproduce neither the UV spectrum nor the optical line profiles.

Nevertheless, an issue of somewhat uncertain implications remains, namely the fact that we cannot at the moment match several observed optical emission lines with a hydrodynamically consistent model that can without difficulties reproduce essentially the entire observable UV range. But with regard to possible discrepancies there is no reason to assume that for modeling O-type atmospheres the optical and the UV spectral ranges carry the same diagnostic weight. The spectral lines in the optical are in general relatively weak and far in-between, compared to the lines in the UV range, and this makes the significance of the UV fit evident when drawing conclusions concerning the quality of the fit in the UV and optical parts of the spectra. Obviously, the parameters from Pauldrach et al. (2004) fit the optical spectrum *and* the UV spectrum, with the exception of H$\alpha$ and He II $\lambda$4686. The currently favored explanation for such discrepancies in the strength of optical emission lines is inhomogeneities in the wind[15] (leading to stronger emission from H and He recombination lines due to the higher densities in the "clumps" compared to a smooth outflow), and an ad-hoc "clumping factor" is often employed to bring the line profiles of the model into agreement with the observed line profiles. But supposing that clumping (or some other effect with similar influence on these two lines) plays a role, then – as long as we have no consistent physical description for this effect – we have complete freedom to fit the optical emission lines without gaining any additional information about the stellar parameters.[16] Thus, a line whose model profile is determined primarily by a cosmetic clumping factor and not the underlying physics of the model completely loses its diagnostic value, and the quality of the fit of this line says nothing about the reliability of the fundamental model parameters.

[14] This result has been verified by two consistently computed models for NGC 6826 and NGC 2392 using the stellar parameter sets of Kudritzki et al. (1997), shown as filled squares in Fig. 17; the abstruse value of the obtained terminal velocity of NGC 6826 is listed in Table 2.

[15] But we note that the amplitudes of the deviations from a smooth, stationary flow are not very large in general (cf. Kudritzki 1999), and we therefore don't expect the clumping to markedly influence the hydrodynamics of the outflow (see, for instance, Pauldrach et al. 1994 and references therein, as well as Runacres & Owocki 2002). This argument is strongly supported by the fact that the hydrodynamic models based on a smooth, time-averaged density structure can indeed reproduce the multitude of UV spectral lines that are formed in the entire atmospheric depth range.

A completely separate avenue of investigation is opened up by an analysis of the behavior of the dynamical wind parameters, independent of the appearance of the spectra. As the wind parameters are not free parameters but (due to the driving mechanism) functions of the stellar parameters, using sets of stellar parameters which are not realized in nature should lead at least partly to an inconsistent (i.e., not observed) behavior of the wind parameters in the analysis. Such an inconsistent behavior can, for example, be evident in the ratio $\upsilon_\infty/\upsilon_{\rm esc}$ of the terminal wind velocity and the escape velocity of individual stars.

We have therefore examined the $\upsilon_\infty/\upsilon_{\rm esc}$ ratios for our CSPN sample as a function of different stellar parameters and compared these to the corresponding ratios from a sample of massive O stars. As already found for massive O stars, the correlation of $\upsilon_\infty$ to $\upsilon_{\rm esc}$ does not represent a strict and simple behavior for CSPNs, giving (as expected) a certain scatter about a mean value. The spread in the $\upsilon_\infty/\upsilon_{\rm esc}$ ratios for our CSPN models turned out to be comparable to that obtained from both massive O star observations and massive O star models, but (as expected from basic physics) at a slightly smaller mean value. Incompatible with all other results, however, are the CSPN "observations", which yield much too high $\upsilon_\infty/\upsilon_{\rm esc}$ ratios in general. These high $\upsilon_\infty/\upsilon_{\rm esc}$ ratios are even more significant because we would have expected the highest $\upsilon_\infty/\upsilon_{\rm esc}$ values to occur for some massive O stars rather than for some CSPNs.

This anomaly is obviously coupled to the fact that the escape velocity $\upsilon_{\rm esc}$ appearing in the "observed" CSPN $\upsilon_\infty/\upsilon_{\rm esc}$ ratios is not a directly observable quantity (whereas $\upsilon_\infty$ can be measured directly). The escape velocity is connected to the stellar mass and radius, and these quantities had been taken from the theoretical post-AGB core-mass–luminosity relation. We thus conclude that there might be a problem with the core-mass–luminosity relation at its high-mass end. This conclusion is further supported by the finding that consistent model calculations using stellar parameters compatible with the core-mass–luminosity relation do yield $\upsilon_\infty/\upsilon_{\rm esc}$ values that lie in the expected range (however, the obtained terminal velocities $\upsilon_\infty$ of these models are highly incompatible with the observations).

Taking these results together, we find that of the three considerations (a) consistency of the dynamical wind parameters with the stellar parameters, (b) compatibility of the model spectra with the observed spectra, and (c) conformity of the stellar parameters with the core-mass–luminosity relation, only two can be reconciled with each other, while the third is irreconcilable with the other two. Since (a) and (b) are strongly supported by both observations and models (spectral appearance, wind-momentum–luminosity relation), it is likely that the consideration that is not realized in nature is (c), the core-mass–luminosity relation, at least for the O-type CSPNs with prominent wind features as studied in this paper.

[16] Note also that without a description of clumping based on first principles, clumping is not a single fit factor but a whole number of them, one for each depth point of the model, and thus the degree of freedom is extremely large. Using these parameters it might be possible to fit the observed optical emission lines even with a wrong hydrodynamical structure.

*Acknowledgements.* We thank Joachim Puls for his advice regarding the implementation of Stark broadening and for running the Fastwind comparison models. We also thank Rolf-Peter Kudritzki for making the optical observations available to us. Thanks to Miguel Urbaneja who shared his original CSPN model runs with us and to an anonymous referee for helpful comments which improved the paper. This work was supported by the Deutsche Forschungsgemeinschaft (DFG) under grant Pa 477/4-1.